\documentclass[pdflatex,sn-mathphys-num,iicol]{sn-jnl}

\usepackage{graphicx}%
\usepackage{multirow}%
\usepackage{amsmath,amssymb,amsfonts}%
\usepackage{amsthm}%
\usepackage{mathrsfs}%
\usepackage[title]{appendix}%
\usepackage{xcolor}%
\usepackage{textcomp}%
\usepackage{manyfoot}%
\usepackage{booktabs}%
\usepackage{algorithm}%
\usepackage{algorithmicx}%
\usepackage{algpseudocode}%
\usepackage{listings}%
\usepackage{bm}
\usepackage{braket}
\usepackage{array}
\usepackage{todonotes}
\usepackage{quantikz}
\usepackage{subfig}

\theoremstyle{thmstyleone}%

\theoremstyle{thmstyletwo}%

\theoremstyle{thmstylethree}%

\begin{document}

\title[Quantum algorithmic framework for finite element analysis]{Quantum algorithmic framework for finite element analysis of isotropic linear elastic materials via efficient block-encoding}

\author*[1]{\fnm{Yuki} \sur{Sato}}\email{yuki-sato@mosk.tytlabs.co.jp}

\author[1]{\fnm{Ruho} \sur{Kondo}}

\affil*[1]{\orgname{Toyota Central R\&D Labs., Inc.}, \orgaddress{\street{1-4-14}, \city{Koraku, Bunkyo-ku, Tokyo}, \postcode{112-0004}, \country{Japan}}}

\abstract{
Quantum computing offers a promising approach to accelerating large-scale numerical analysis, including the finite element method. 
However, its practical application requires efficient quantum circuit implementations of discretized linear systems. 
In this study, we develop a quantum algorithmic framework for two-dimensional isotropic linear elastic materials discretized using the finite element method.
The key contribution is an explicit construction of a normalized block-encoding of the shifted global stiffness matrix, which is essential for applying a quantum linear system solver based on the quantum singular value transformation with improved dependence on the condition number.
By analytically diagonalizing the element stiffness matrix and expressing the global stiffness matrix through a factorized finite element assembly, we construct the required block-encoding using a constant number of ancilla qubits and a gate complexity logarithmic in the number of degrees of freedom. 
We further analyze the computational complexity of displacement state preparation, strain energy estimation, and low-resolution displacement reconstruction, identifying conditions under which polynomial quantum speedups may be achieved. 
Numerical experiments on cantilever beam problems verify the accuracy of the proposed quantum circuits and demonstrate agreement with conventional finite element solutions. 
These results establish an explicit and resource-efficient quantum framework for finite element analysis of isotropic linear elastic materials.
}


\maketitle

\section{Introduction}\label{sec:introduction}

Computational mechanics is a fundamental discipline that provides numerical methodologies for analyzing and predicting the behavior of physical systems governed by partial differential equations (PDEs). 
Among the various numerical approaches, the finite element method (FEM) has become one of the most widely adopted techniques owing to its flexibility in handling complex geometries, heterogeneous materials, and multiphysics problems~\cite{hughes2000finite,zienkiewicz2013finite}. 
Despite its remarkable success, the computational cost of FEM can be significantly large, particularly for large-scale simulations and high-fidelity analyses. 
As scientific and engineering applications continue to demand higher resolution and more accurate simulations, developing computational paradigms that have better scalability has become an increasingly important research topic.

Quantum computing has recently emerged as a promising computational paradigm capable of solving certain classes of computational problems more efficiently than classical computers. 
Quantum algorithms have been investigated for various problems in computational mechanics, including wave equations~\cite{costa2019quantum, sato2024hamiltonian}, heat equations~\cite{sato2025quantum, jin2026schr}, electromagnetic wave equations~\cite{jin2024quantum, tezuka2026quantum}, and elastic wave equation~\cite{bosch2025quantum, nakanishi2026hamiltonian}.
Within this broader effort, several quantum finite element methods have been proposed by integrating quantum linear system algorithms, including the Harrow–Hassidim–Lloyd (HHL) algorithm~\cite{harrow2009quantum} and its subsequent developments~\cite{childs2017quantum, an2022quantum}, into conventional finite element formulations. 
Montanaro and Pallister~\cite{montanaro2016quantum} performed a complexity analysis of quantum FEM based on the quantum linear system solver~\cite{childs2017quantum} and showed that, for fixed spatial dimension, the expected advantage is generally polynomial with a more favorable separation in higher dimensions. 
Qu-FEM~\cite{alkadri2025quantum} introduced local interaction primitives and a local-to-global indicator matrix to block-encode the stiffness matrix for a modified Poisson problem and performed the finite element analysis based on the adiabatic quantum computing-based linear system solver~\cite{an2022quantum}.
Perera et al.~\cite{perera2026iqfem} proposed an immersed quantum finite element method (IQFEM) for Poisson problems with heterogeneous coefficients and complex geometries where they constructed a block-encoding with polylogarithmic gate complexity in the number of degrees of freedom.
Danz et al.~\cite{danz2026implementation} constructed explicit quantum oracles for stiffness and mass matrices arising in finite element analysis of elastic structures and demonstrated polylogarithmic scaling with the number of degrees of freedom under suitable assumptions.
Furthermore, Suresh~\cite{suresh2026quantum} constructed block-encodings of two-dimensional finite element operators for Poisson and linear elasticity problems under periodic boundary conditions, using a linear combination of unitaries (LCU) with a constant number of terms.
While these studies have advanced the efficient quantum representation of finite element operators, conventional quantum linear system solvers generally exhibit a query complexity with at least linear dependence on the condition number $\kappa$. 
In contrast, it is well known that classical iterative methods for symmetric positive-definite matrices, such as the conjugate gradient method, achieve a square-root dependence on $\kappa$~\cite{saad2003iterative}. 
Orsucci and Dunjko~\cite{orsucci2021solving} have established a similar improvement for quantum linear system solvers, showing that the query complexity can be reduced to $\widetilde{O}(\sqrt{\kappa})$ for positive-definite matrices. 
However, this improvement requires an efficient normalized block-encoding of a shifted coefficient matrix, which is not directly guaranteed by an efficient block-encoding of the original coefficient matrix.

This work addresses this requirement for two-dimensional isotropic linear elastic materials discretized by bilinear $Q_1$ elements on a rectangular mesh. 
We analytically diagonalize the element stiffness matrix and express the global stiffness matrix through a factorized finite element assembly.
Then, we explicitly construct a normalized block-encoding of the shifted stiffness matrix with a gate count logarithmic in the number of degrees of freedom, up to the costs of implementing oracles for material distribution, boundary conditions, and distributed loads.
We further analyze the computational complexity of displacement state preparation, strain energy estimation, and low-resolution displacement reconstruction, identifying conditions under which polynomial quantum speedups may be achieved.

The remainder of this paper is organized as follows. Section~\ref{sec:preliminary} introduces the finite element formulation for isotropic linear elastic materials and the quantum algorithmic preliminaries, including quantum singular value transformation (QSVT) and low-degree approximation of inverse operators. 
Section~\ref{sec:method} presents the explicit construction of the normalized block-encoding of the shifted stiffness matrix and analyzes the computational complexity of displacement state preparation, strain energy estimation, and low-resolution displacement reconstruction. 
Section~\ref{sec:results} provides numerical experiments to validate the proposed quantum circuit implementation. 
Section~\ref{sec:conclusions} concludes this study and discusses future research directions.

\section{Preliminaries}\label{sec:preliminary}

\subsection{Finite element method for isotropic linear elastic materials}

First, we briefly explain the finite element method for isotropic linear elastic material.
Let $\Omega \subset \mathbb{R}^d~(d=2,3)$ be a $d$-dimensional open bounded domain occupied by a linear elastic material.
For linear elastic materials, the constitutive relation between the stress tensor $\boldsymbol{\sigma}$ and the infinitesimal strain tensor $\boldsymbol{\varepsilon}$ is given by $\boldsymbol{\sigma} = \boldsymbol{C} : \boldsymbol{\varepsilon}$, where $\boldsymbol{C}$ denotes the fourth-order elasticity tensor. 
The infinitesimal strain tensor is defined as $\boldsymbol{\varepsilon}=(\nabla \boldsymbol{u} + \nabla \boldsymbol{u}^{\top})/2$, where $\boldsymbol{u}$ is the displacement field and $\nabla$ is the spatial gradient operator.
For isotropic linear elastic materials, the elasticity tensor is characterized by the two Lam\'{e} parameters, $\lambda$ and $\mu$, and the constitutive equation can be written as 
\begin{equation}
    \boldsymbol{\sigma} = 2 \mu \boldsymbol{\varepsilon} + \lambda \,\mathrm{tr}(\boldsymbol{\varepsilon}) \boldsymbol{I},
\end{equation}
where $\mathrm{tr}(\cdot)$ denotes the trace operator and $\boldsymbol{I}$ is the identity tensor.
The Lam\'{e} parameters are related to Young's modulus $E$ and Poisson's ratio $\nu$ by 
\begin{align}
    \lambda &= \frac{E\nu}{(1+\nu)(1-2\nu)} \\
    \mu &= \frac{E}{2(1+\nu)}.
\end{align}

For two-dimensional problems, two commonly used assumptions are the plane strain and plane stress conditions. 
Under the plane strain assumption, the constitutive relation is obtained directly from the three-dimensional formulation.
In contrast, under the plane stress assumption, eliminating the out-of-plane stress results in an effective Lam\'{e} parameter $\lambda_\mathrm{eff}=E\nu/(1-\nu^{2})$ while the other parameter $\mu$ remains unchanged.

The displacement field is governed by the following equilibrium equation:
\begin{equation}
    -\nabla\cdot\boldsymbol{\sigma}=\boldsymbol{f}
\qquad\text{in }\Omega,    
\end{equation}
where $\boldsymbol{f}$ is the body force applied to the material.
The boundary of the domain $\Omega$ is partitioned into a Dirichlet boundary $\Gamma_\mathrm{D}$ and a Neumann boundary $\Gamma_\mathrm{N}$, that is, $\partial \Omega = \Gamma_\mathrm{D} \cup \Gamma_\mathrm{N}$ and $\Gamma_\mathrm{D} \cap \Gamma_\mathrm{N} = \emptyset$. 
Then, the boundary conditions are prescribed as
\begin{align}
    \boldsymbol{u} &= \boldsymbol{g} \quad \text{on } \Gamma_\mathrm{D}, \\
    \boldsymbol{\sigma} \boldsymbol{n} &= \boldsymbol{t} \quad \text{on } \Gamma_\mathrm{N},
\end{align}
where $\boldsymbol{n}$ is the outward unit normal vector, $\boldsymbol{g}$ is the prescribed displacement, and $\boldsymbol{t}$ is the prescribed surface traction.

To solve the above boundary value problem by the finite element method, we consider the weak form of the governing equation.
Let $\mathcal{U}$ and $\mathcal{V}$ be the admissible and test spaces defined, respectively, as 
\begin{align}
    \mathcal{U} := \{ \boldsymbol{v} \in H^1(\Omega)^d ~|~ \boldsymbol{v} = \boldsymbol{g} \text{ on } \Gamma_\mathrm{D} \} \\
    \mathcal{V} := \{ \boldsymbol{v} \in H^1(\Omega)^d ~|~ \boldsymbol{v} = \boldsymbol{0} \text{ on }\Gamma_\mathrm{D} \}
\end{align}
where $H^1(\Omega)$ is the Sobolev space defined on $\Omega$.
Multiplying the equilibrium equation by a test function $\boldsymbol{v} \in \mathcal{V}$, integrating over the computational domain $\Omega$, and applying the divergence theorem yield the weak formulation: 
find $\boldsymbol{u} \in \mathcal{U}$ such that
\begin{align}
    a(\boldsymbol{u}, \boldsymbol{v}) = \ell(\boldsymbol{v})
    \qquad \forall \boldsymbol{v} \in \mathcal{V},
    \label{eq:weak-form}
\end{align}
with
\begin{align}
    a(\boldsymbol{u},\boldsymbol{v})
    &= \int_\Omega \boldsymbol{\varepsilon}(\boldsymbol{v}) :
    \boldsymbol{C} : \boldsymbol{\varepsilon}(\boldsymbol{u})\,\mathrm{d}\Omega,
    \label{eq:bilinear-form}\\
    \ell(\boldsymbol{v})
    &=\int_\Omega \boldsymbol{v} \cdot \boldsymbol{f} \,\mathrm{d}\Omega
    + \int_{\Gamma_N} \boldsymbol{v} \cdot \boldsymbol{t}\,\mathrm{d}\Gamma.
    \label{eq:linear-form}
\end{align}
Assuming that the Dirichlet boundary $\Gamma_\mathrm{D}$ has a positive measure, the bilinear form $a(\cdot,\cdot)$ is continuous and coercive on $\mathcal{V}$ by Korn's inequality.
Moreover, the linear functional $\ell(\cdot)$ is bounded on $\mathcal{V}$.
Therefore, by the Lax--Milgram theorem, the weak problem \eqref{eq:weak-form} admits a unique solution $\boldsymbol{u} \in \mathcal{U}$.

Next, we discretize the weak problem \eqref{eq:weak-form} by the finite element method.
Let $\mathcal{T}_h$ be a conforming partition of the domain $\Omega$ into finite elements.
The finite-dimensional trial and test spaces are then defined as
\begin{align}
    \mathcal{U}_h \subset \mathcal{U}, \qquad
    \mathcal{V}_h \subset \mathcal{V}.
\end{align}
The weak problem is now approximated as the problem for finding $\boldsymbol{u}_h \in \mathcal{U}_h$ satisfying
\begin{align}
    a(\boldsymbol{u}_h,\boldsymbol{v}_h)
    =
    \ell(\boldsymbol{v}_h),
    \qquad
    \forall
    \boldsymbol{v}_h\in\mathcal{V}_h.
\end{align}
To solve this problem, let $\{\phi_j\}_{j=1}^{N_\mathrm{ve}}$ denote the nodal basis functions, where $N_\mathrm{ve}$ is the number of nodes of finite elements.
The displacement field is approximated as
\begin{align}
    \boldsymbol{u}_h(\boldsymbol{x})
    =
    \sum_{j=1}^{N}
    \phi_j(\boldsymbol{x})
    \boldsymbol{u}_j,
\end{align}
where $\boldsymbol{u}_j \in \mathbb{R}^{d}$ is the displacement vector associated with the $j$-th node.
In the present work, we employ piecewise bilinear $Q_1$ basis functions.
Since each nodal basis function has local support, it is nonzero only on the elements sharing the corresponding node.
Therefore, the domain integral appearing in the bilinear form \eqref{eq:bilinear-form} can be decomposed into a sum of element-wise integrals, that is,
\begin{align}
    a(\boldsymbol{u}_h,\boldsymbol{v}_h)
    =
    \sum_{e=1}^{N_\mathrm{el}}
    \int_{\Omega_e}
    \boldsymbol{\varepsilon}(\boldsymbol{v}_h)
    :
    \boldsymbol{C}
    :
    \boldsymbol{\varepsilon}(\boldsymbol{u}_h)
    \,\mathrm{d}\Omega,
\end{align}
where $\Omega_e$ denotes the domain of the $e$-th element and $N_\mathrm{el}$ is the total number of elements.
This element-wise decomposition motivates introducing the element stiffness matrix.
For each element $e\in\mathcal{T}_h$, the element stiffness matrix is defined by
\begin{align}
    \mathbf{K}_e
    =
    \int_{\Omega_e}
    \mathbf{B}^{\top}
    \mathbf{C}
    \mathbf{B}
    \,
    \mathrm{d}\Omega,
    \label{eq:element-stiffness}
\end{align}
where $\Omega_e$ denotes the element domain and $\mathbf{B}$ is the strain--displacement matrix depending on the nodal basis functions and $\mathbf{C}$ is the matrix representation (Voigt notation) of the elastic tensor.
The global stiffness matrix is then obtained by assembling the element stiffness matrices over all elements,
\begin{align}
    \mathbf{K}
    =
    \sum_{e=1}^{N_\mathrm{el}}
    (\mathbf{R}_e)^{\top}
    \mathbf{K}_e
    \mathbf{R}_e,
    \label{eq:global-stiffness}
\end{align}
where $N_\mathrm{el}$ is the number of elements and
$\mathbf{R}_e$ is the Boolean element restriction matrix that maps the global degrees of freedom to the local degrees of freedom of element $e$.
The resulting finite element system is given by
\begin{align}
    \mathbf{K}\mathbf{u}
    =
    \mathbf{f},
    \label{eq:fem-system}
\end{align}
where $\mathbf{u}$ and $\mathbf{f}$ denote the global displacement and load vectors, respectively.
Equation \eqref{eq:global-stiffness} is the standard finite element assembly formula. 
In Section \ref{subsec:global-block-encoding}, we reformulate this assembly process into a tensor-product representation that is particularly suitable for constructing an efficient block-encoding of the global stiffness matrix.

\subsection{Quantum computing}
\label{subsec:quantum-preliminaries}

\subsubsection{Quantum states}
\label{subsubsec:quantum-notation}

Here, we introduce the basic concept of quantum computing needed for the quantum linear system algorithm used in this work.
Quantum computing uses a qubit (quantum bit) as the fundamental unit that can exist in a superposition of the states 0 and 1, unlike a classical bit that can only be either 0 or 1.
A qubit has two computational basis states, defined as
\begin{equation}
    \ket{0}:=\begin{pmatrix}1\\0\end{pmatrix},
    \qquad
    \ket{1}:=\begin{pmatrix}0\\1\end{pmatrix},
\end{equation}
and a general pure state is expressed as $c_0\ket{0}+c_1\ket{1}$, where $c_0,c_1\in\mathbb{C}$ and $|c_0|^2+|c_1|^2=1$. 
Quantum measurement in the computational basis returns $0$ or $1$ with probabilities $|c_0|^2$ and $|c_1|^2$, respectively.

An $n$-qubit register spans $\mathcal{H}_n \cong \mathbb{C}^{2^n}$, with computational basis states $\ket{q_{n-1}, q_{n-2}, \ldots, q_0}:=\ket{q_{n-1}} \otimes \ket{q_{n-2}} \otimes \cdots \otimes \ket{q_0}$ where $q_j \in \{0, 1\}$.
In the conventional notation, we express $\ket{q} := \ket{q_{n-1}, q_{n-2}, \ldots, q_0}$ for $q \in \{0, \ldots, 2^n-1\}$ where $q := \sum_{j=0}^{n-1} q_j 2^j = (q_{n-1}, q_{n-2}, \ldots, q_0)_2$. 
That is, the computational basis state $\ket{q}$ is the $2^n$-dimensional vector whose $q$-th entry is one and all other entries are zero, using zero-based indexing.

For a nonzero vector $\mathbf{v} \in \mathbb{C}^N$, its amplitude encoding is given as
\begin{equation}
    \ket{v} = \frac{1}{ \| \mathbf{v} \|_2 }
    \sum_{q=0}^{N-1} v_q \ket{q},
    \label{eq:amplitude-encoding}
\end{equation}
using $n = \lceil \log_2 N \rceil$ qubits.
When $N<2^n$, the amplitudes of the unused basis states are set to zero.
Note that the vector $\mathbf{v}$ must be normalized so that it can be encoded on a quantum state.
A measurement of $\ket{v}$ returns an index $q$ with probability
$|v_q|^2/\|v\|_2^2$.

\subsubsection{Block-encoding}
\label{subsubsec:block-encoding}

A unitary $\mathbf{U}_A$ acting on an $n$-qubit system register and $a$ ancilla qubits is called an $(\alpha, a, \epsilon)$ block-encoding of $\mathbf{A} \in \mathbb{C}^{2^n \times 2^n}$ if $\mathbf{U}_A$ satisfies
\begin{equation}
    \|
    \mathbf{A} - \alpha
    \left(\bra{0^a} \otimes \mathbf{I}^{(n)} \right)
    \mathbf{U}_A
    \left(\ket{0^a} \otimes \mathbf{I}^{(n)} \right)\|
    \leq \epsilon,
    \label{eq:block-encoding-definition}
\end{equation}
where $\mathbf{I}^{(n)}$ is the identity matrix with size of $2^n \times 2^n$ and $\ket{0^a}$ is the $a$-qubit zero state.
Since the construction in this paper is algorithmically exact, we regard $\epsilon = 0$ unless stated otherwise. 
The scalar $\alpha \geq \| \mathbf{A} \|$ is the block-encoding normalization or subnormalization factor. 
The block-encoding can also be represented as
\begin{align}
    \mathbf{U}_A =
    \begin{pmatrix}
        \mathbf{A} / \alpha & \ast \\
        \ast & \ast
    \end{pmatrix},
\end{align}
where $\ast$ denotes a block whose explicit form is not relevant here.
The normalization (or subnormalization) factor $\alpha$ is introduced to enable an arbitrary matrix $\mathbf{A}$ to be embedded into a block of a unitary matrix.

By applying $\mathbf{U}_A$ to a normalized input $\ket{v}$, we obtain
\begin{equation}
    \mathbf{U}_A \left( \ket{0^a}\otimes\ket{v} \right)
    =
    \ket{0^a}\otimes\frac{\mathbf{A}\ket{v}}{\alpha}
    +\ket{\perp},
\end{equation}
where $\ket{\perp}$ is the state orthogonal to the first term, specifically, $(\bra{0^a}\otimes \mathbf{I}^{(n)}) \ket{\perp}=0$.
If $\mathbf{A}\ket{v}\neq 0$, the postselection on the all-zero outcome for ancillas produces the normalized state $\mathbf{A}\ket{v} / \|\mathbf{A} \ket{v}\|$, with probability $\|\mathbf{A} \ket{v}\|^2 / \alpha^2$.

\subsubsection{Quantum singular value transformation}
\label{subsubsec:qsvt-preliminaries}

Let $\mathbf{U}_A$ be a block-encoding of a matrix $\mathbf{A}$ with normalization factor $\alpha$, and let $\mathbf{\Pi}$ denote the projector onto the encoded subspace.
Quantum singular value transformation (QSVT)~\cite{gilyen2019qsvt, martyn2021grand} is a technique for implementing polynomial transformations of the singular values of a block-encoded matrix by alternating applications of $\mathbf{U}_A$, its adjoint $\mathbf{U}_A^\dagger$, and projector-controlled phase-shift operators defined as
\begin{equation}
    \mathbf{\Pi}_{\phi} := e^{i \phi (2 \mathbf{\Pi} - \mathbf{I})},
    \label{eq:projector-phase}
\end{equation}
which can be implemented using phase rotations controlled by the ancilla register of the block-encoding.
Specifically, a sequence of alternating applications of $\mathbf{U}_A$, $\mathbf{U}_A^\dagger$, and $\mathbf{\Pi}_\phi$ yields a unitary $\mathbf{U}_{\vec{\phi}}$ defined, for an odd constant $m$, as
\begin{align}
    \mathbf{U}_{\vec{\phi}} &:= \mathbf{\Pi}_{\phi_1} \mathbf{U}_A \prod_{k=1}^{(m-1)/2} \left( \mathbf{\Pi}_{\phi_{2k}} \mathbf{U}_A^\dagger \mathbf{\Pi}_{\phi_{2k+1}} \mathbf{U}_A \right) \nonumber \\
    &= \begin{pmatrix}
        P \left( \mathbf{A} / \alpha \right) & \ast \\
        \ast & \ast
    \end{pmatrix},
\end{align}
and for an even $m$,
\begin{align}
    \mathbf{U}_{\vec{\phi}} &:= \prod_{k=1}^{m/2} \left( \mathbf{\Pi}_{\phi_{2k-1}} \mathbf{U}_A^\dagger \mathbf{\Pi}_{\phi_{2k}} \mathbf{U}_A \right) \nonumber \\
    &= \begin{pmatrix}
        P \left( \mathbf{A} / \alpha \right) & \ast \\
        \ast & \ast
    \end{pmatrix},
\end{align}
where $P(x)$ is a polynomial, and $P \left( \mathbf{A} / \alpha \right)$ means the polynomial transformation of the singular value of $\mathbf{A} / \alpha$.
For an admissible real polynomial $P(x)$, the phases $\{\phi_k\}_{k=0}^{m}$ can be determined by solving an angle finding problem for quantum signal processing~\cite{dong2021efficient, dong2024robust}. 
The polynomial must satisfy (i) $\deg(P)\le m$, (ii) $P(-x)=(-1)^m P(x)$, and (iii) $|P(x)|\le 1$ for all $x\in[-1,1]$. 
Thus, $P(x)$ must be odd for odd $m$ and even for even $m$.
For a Hermitian positive semidefinite matrix, this singular value transformation reduces to a polynomial eigenvalue transformation.
The QSVT requires exactly $m$ queries to $\mathbf{U}_A$ and $\mathbf{U}_A^\dagger$ in total, together with $O(m)$ projector-controlled phase shifts. 
Consequently, the query complexity of implementing a polynomial transformation scales linearly with its degree.

\subsubsection{QSVT-based quantum linear system solvers}
In this study, we focus on solving Eq.~\eqref{eq:fem-system}.
To this end, we employ QSVT, which enables the polynomial transformations of the singular values of a block-encoded matrix. 
By constructing a degree-$m$ polynomial that approximates the reciprocal function, we implement the pseudoinverse of the block-encoded matrix.
Since a polynomial implemented by QSVT must have magnitude at most one over the interval $[-1, 1]$, the target function is scaled appropriately so that it satisfies this boundedness requirement.
Let
\begin{align}
    \gamma := \frac{\| \mathbf{A} \|}{\alpha} \leq 1.
\end{align}
Then, the singular values of the normalized matrix $\mathbf{A}/\alpha$ lie in the interval $[\gamma/\kappa, \gamma]$, where $\kappa$ is the condition number of $\mathbf{A}$. 
Accordingly, the polynomial is required to approximate the scaled reciprocal function $\gamma / (\kappa x)$ over the domain $[-\gamma ,-\gamma/\kappa] \cup [\gamma/\kappa, \gamma]$.
It is known that such a polynomial can be constructed with degree $m \in \mathcal{O}((\kappa/\gamma) \log (\kappa / \epsilon) )$ \cite{sunderhauf2025matrix}, yielding a QSVT-based quantum linear system algorithm with the asymptotic query complexity $\mathcal{O}((\kappa/\gamma) \log (\kappa / \epsilon) )$ where $\epsilon$ is the target approximation error.

\subsubsection{Low degree approximation of inverse operators}
\label{subsubsec:orsucci-dunjko}

The polynomial for approximating the reciprocal function described above has a degree that is essentially linear in the condition number when it is applied directly to a block-encoding of the coefficient matrix.
Orsucci and Dunjko showed that, for problems admitting an efficient \emph{normalized} block-encoding of a shifted matrix, the inverse matrix can instead be implemented with a polynomial degree proportional to the square root of a condition number~\cite{orsucci2021solving}.
We summarize the part of their construction used in this work.

Let $\mathbf{A}\succeq 0$ be Hermitian, and $\varsigma_\mathrm{min}^{+}(\mathbf{A})$ denote its smallest nonzero eigenvalue.
Using $\beta > 0$ such that $\beta\|\mathbf{A}\|\leq 2$, we define
\begin{equation}
    \mathbf{B}:=\mathbf{I} - \beta \mathbf{A}.
    \label{eq:OD-shifted-matrix}
\end{equation}
Suppose that a normalized $(1,a_B,0)$-block-encoding $\mathbf{U}_B$ of $\mathbf{B}$ is available, and that a constant $\kappa_B > 1$ satisfies
\begin{equation}
    \beta \varsigma_\mathrm{min}^{+}(\mathbf{A})
    \geq \frac{1}{\kappa_B}.
    \label{eq:OD-condition}
\end{equation}
The eigenvalues of $\mathbf{B}$ corresponding to the non-zero eigenvalues of $\mathbf{A}$ then lie in
\begin{equation}
    \mathcal{D}_B
    :=\left[-1,1-\frac{1}{\kappa_B}\right].
    \label{eq:OD-B-domain}
\end{equation}
If $\mathbf{A}$ is singular, its zero-eigenvalue eigenspace corresponds to the eigenspace of $\mathbf{B}$ associated with the eigenvalue one.

Our goal is to construct a polynomial that approximates $1/(1-x)$ on $\mathcal{D}_B$ while vanishing at $x=1$.
As shown later, applying the singular value transformation with this polynomial to $\mathbf{B}$ yields the Moore--Penrose pseudoinverse of $\mathbf{A}$.
To this end, we consider a polynomial of the form $[1-\widehat{T}(x)]^2/(1-x)$, where $\widehat{T}(x)$ is chosen to be close to zero on $\mathcal{D}_B$ and to satisfy $\widehat{T}(1)=1$.
Let $T_{\ell}$ denote the degree-$\ell$ Chebyshev polynomial of the first kind. 
The shifted and normalized polynomial is defined as follows.
\begin{equation}
    \widehat{T}_{\ell,\kappa_B}(x)
    :=
    \frac{
    T_{\ell}\!\left(
    \dfrac{x+1/(2\kappa_B)}{1-1/(2\kappa_B)}
    \right)
    }{
    T_{\ell}\!\left(
    \dfrac{1+1/(2\kappa_B)}{1-1/(2\kappa_B)}
    \right)
    }.
    \label{eq:OD-shifted-Chebyshev}
\end{equation}
The affine transformation in the argument of $T_\ell$ maps $\mathcal{D}_B$ onto $[-1,1]$, where $|T_\ell(x)|\leq 1$.
Meanwhile, the denominator normalizes the polynomial
such that $\widehat{T}_{\ell,\kappa_B}(1)=1$.
Since the argument of $T_\ell$ in the denominator is greater than one, the denominator grows exponentially with respect to $\ell/\sqrt{\kappa_B}$.
Consequently, $\widehat{T}_{\ell,\kappa_B}(x)$
converges exponentially to zero on $\mathcal{D}_B$ as $\ell$ increases.

Then, the polynomial for inversion is defined as
\begin{equation}
    P_{2\ell-1,\kappa_B}(x)
    :=
    \frac{
    \left[1-\widehat{T}_{\ell,\kappa_B}(x)\right]^2
    }{1-x}.
    \label{eq:OD-inverse-polynomial}
\end{equation}
Although Eq.~\eqref{eq:OD-inverse-polynomial} is written as a quotient, the numerator has a second-order zero at $x=1$.
Hence, the quotient is a polynomial of degree $2\ell-1$ and satisfies
\begin{equation}
    P_{2\ell-1,\kappa_B}(1)=0.
    \label{eq:OD-kernel-zero}
\end{equation}
This condition ensures that the kernel of $\mathbf{A}$, corresponding to the eigenspace of $\mathbf{B}$ associated with the eigenvalue one, is mapped to zero, as required by the Moore--Penrose pseudoinverse.
For
\begin{equation}
    \ell\in
    \Theta\!\left(
    \sqrt{\kappa_B}
    \log\frac{\kappa_B}{\epsilon_{\mathrm{poly}}}
    \right),
    \label{eq:OD-polynomial-degree}
\end{equation}
the polynomial approximates the shifted reciprocal function on $\mathcal{D}_B$ as
\begin{equation}
    \max_{x\in\mathcal{D}_B}
    \left|
    P_{2\ell-1,\kappa_B}(x)-\frac{1}{1-x}
    \right|
    \leq \epsilon_{\mathrm{poly}}.
    \label{eq:OD-polynomial-error}
\end{equation}
To see how this approximation yields the pseudoinverse, consider a nonzero eigenvalue $\varsigma$ of $\mathbf{A}$.
The corresponding eigenvalue of $\mathbf{B}$ is $x=1-\beta\varsigma$, and hence
\begin{equation}
    P_{2\ell-1,\kappa_B}(1-\beta\varsigma)
    \approx \frac{1}{\beta\varsigma}.
\end{equation}
Therefore, $\beta P_{2\ell-1,\kappa_B}(\mathbf{B})$ approximates $\mathbf{A}^{+}$ on its nonzero-eigenvalue subspace, while Eq.~\eqref{eq:OD-kernel-zero} ensures that its kernel is mapped to zero.
In particular,
\begin{equation}
    \left\|
    \beta P_{2\ell-1,\kappa_B}(\mathbf{B})
    -\mathbf{A}^{+}
    \right\|
    \leq\beta\epsilon_{\mathrm{poly}}.
\end{equation}

To implement this polynomial transformation using QSVT, the polynomial must be bounded in magnitude by one over $[-1,1]$.
We therefore define
\begin{align}
    P_\mathrm{max}
    &:=2\max_{x\in[-1,1]}
    \left|P_{2\ell-1,\kappa_B}(x)\right|, \\
    \widehat{P}(x)
    &:=\frac{P_{2\ell-1,\kappa_B}(x)}{P_\mathrm{max}}.
    \label{eq:OD-normalized-polynomial}
\end{align}
This normalization ensures $|\widehat{P}(x)|\leq 1/2$ for all $x\in[-1,1]$.
For the degree in Eq.~\eqref{eq:OD-polynomial-degree}, $P_\mathrm{max}\in\Theta(\kappa_B)$~\cite{orsucci2021solving}.
The polynomial in Eq.~\eqref{eq:OD-inverse-polynomial} generally contains both even and odd polynomials.
In our implementation, it is decomposed as
\begin{equation}
    \widehat{P}
    =\widehat{P}_{\mathrm{even}}
    +\widehat{P}_{\mathrm{odd}}.
\end{equation}
Since $|\widehat{P}(x)|\leq 1/2$ on $[-1,1]$, both $2\widehat{P}_{\mathrm{even}}$ and
$2\widehat{P}_{\mathrm{odd}}$ are bounded in magnitude by one on this interval.
Thus, these are implemented by separate even- and odd-parity QSVT and combined by a linear combination of unitaries (LCU).
Consequently, the top-left block of the resulting unitary implements $\widehat{P}(\mathbf{B})$, which satisfies
\begin{equation}
    \left\|
    \widehat{P}(\mathbf{B})
    -\frac{\mathbf{A}^{+}}{\beta P_\mathrm{max}}
    \right\|
    \leq
    \frac{\epsilon_{\mathrm{poly}}}{P_\mathrm{max}}.
    \label{eq:OD-pseudoinverse-block}
\end{equation}
The number of calls to $\mathbf{U}_B$ and $\mathbf{U}_B^{\dagger}$ required to construct this pseudoinverse block-encoding is
\begin{equation}
    \mathcal{O}\!\left(
    \sqrt{\kappa_B}
    \log\frac{\kappa_B}{\epsilon_{\mathrm{poly}}}
    \right).
    \label{eq:OD-query-complexity}
\end{equation}
Note that preparing the normalized solution state additionally requires postselection or amplitude amplification, whose cost depends on the success amplitude $\|\mathbf{A}^{+}\ket{b}\|/(\beta P_\mathrm{max})$~\cite{orsucci2021solving}.

\section{Method} \label{sec:method}

As described in Section~\ref{subsubsec:orsucci-dunjko}, the low-degree polynomial approximation of the inverse operator requires a normalized block-encoding of $\mathbf{B} := \mathbf{I} - \beta \mathbf{K}$, where $\mathbf{K}$ denotes the global stiffness matrix.
Hence, the central objective of our construction is to obtain an efficient normalized block-encoding of the shifted stiffness matrix.
It is known that an efficient block-encoding of $\mathbf{K}$
does not necessarily provide the normalized block-encoding of $\mathbf{B}$ required by the low-degree approximation of inverse operators~\cite{orsucci2021solving}.
Here, we implement the normalized block-encoding by exploiting the structure of finite element assembly
with an analytic diagonalization of the element stiffness matrix.

\subsection{Factorized representation of the global stiffness matrix}
\label{subsec:global-block-encoding}

We consider an $N_\mathrm{x} \times N_\mathrm{y}$ structured two-dimensional mesh consisting of square elements.
The number of nodes is
\begin{equation}
    N_\mathrm{ve} = (N_\mathrm{x}+1)(N_\mathrm{y}+1),
\end{equation}
and the number of degrees of freedom is
\begin{equation}
    N = 2 N_\mathrm{ve} = 2(N_\mathrm{x}+1)(N_\mathrm{y}+1).
    \label{eq:number-dofs}
\end{equation}
The global nodes are labeled by the zero-based indices $(v_\mathrm{x}, v_\mathrm{y})$, where $v_\mathrm{x} \in \{0, \ldots, N_\mathrm{x} \}
$ and $v_\mathrm{y} \in \{0, \ldots, N_\mathrm{y} \}$ denote the node indices in the $x$- and $y$-directions, respectively, and the displacement component is indexed by $ v_\mathrm{c} \in \{0,1\}$ where $v_\mathrm{c}=0$ represents the $x$-component and $v_\mathrm{c}=1$ the $y$-component.
The number of qubits required for representing node indices are $n_\mathrm{x} + n_\mathrm{y}$ where $n_\mathrm{x} := \lceil \log_2 (N_\mathrm{x} + 1) \rceil$ and $n_\mathrm{y} := \lceil \log_2 (N_\mathrm{y} + 1) \rceil$, and one qubit is used to represent the displacement component $v_\mathrm{c}$.
The number of elements is 
\begin{equation}
    N_\mathrm{el} = N_\mathrm{x} N_\mathrm{y}.
\end{equation}
Each element is labeled by the zero-based indices $(e_\mathrm{x}, e_\mathrm{y})$, where $e_\mathrm{x} \in \{0, \ldots, N_\mathrm{x} - 1 \}$ and $e_\mathrm{y} \in \{0, \ldots, N_\mathrm{y} - 1 \}$ denote the element indices in the $x$- and $y$-directions, respectively.
Four local nodes of an element are labeled by $(v_\xi, v_\eta) \in \{0,1\}^2$.

Let $\mathbf{K}_e \in \mathbb{R}^{8\times 8}$ denote the element stiffness matrix, where the dimension $8$ arises from the four nodes of a quadrilateral element, each having two displacement degrees of freedom in the $x$- and $y$-directions.
Let $\mathbf{R}_e\in\{0,1\}^{8\times N}$ be an element restriction operator that acts as
\begin{equation}
    \boldsymbol{u}_e=\mathbf{R}_e\boldsymbol{u},
\end{equation}
where $\boldsymbol{u}_e$ is the local displacement vector of the element $e$.
In conventional FEM, the contributions $\mathbf{R}_e^\top\mathbf{K}_e\mathbf{R}_e$ can be accumulated directly into a global stiffness matrix. 
However, in quantum computing, the stiffness matrix, which is generally nonunitary, must be embedded into a larger unitary through block-encoding, whose normalization affects the cost of the subsequent inverse operation. 
For example, a straightforward construction of Eq.~\eqref{eq:global-stiffness} by LCU may introduce a subnormalization that grows with the number of elements. 
This motivates rewriting the assembly as a product of matrices, which is suitable for quantum computing.
Stacking all restriction matrices gives
\begin{equation}
    \mathbf{R}=
    \begin{bmatrix}
        \mathbf{R}_1^{\top} & \cdots & \mathbf{R}_{N_\mathrm{el}}^{\top}
    \end{bmatrix}^{\top}
    \in \{0 , 1 \}^{8N_\mathrm{el} \times N}.
    \label{eq:stacked-S}
\end{equation}
The global stiffness matrix by the finite element assembly is then represented as
\begin{equation}
    \mathbf{K} = \mathbf{R}^{\top}
    \left(\bigoplus_{e=1}^{N_\mathrm{el}}\mathbf{K}_e\right) \mathbf{R}.
    \label{eq:assembled-general}
\end{equation}
For a homogeneous material on the uniform square mesh, Eq.~\eqref{eq:assembled-general} reduces to
\begin{equation}
    \mathbf{K}=\mathbf{R}^{\top}
    \left(\mathbf{I}^{(N_\mathrm{el})} \otimes \mathbf{K}_e \right)\mathbf{R}.
    \label{eq:assembled-homogeneous}
\end{equation}
This factorized representation of finite element assembly enables us to obtain a normalized Gram factor $\mathbf{L}$ as
\begin{align}
    \mathbf{L} := \frac{1}{\sqrt{\alpha_K}} \left(\mathbf{I}^{(N_\mathrm{el})} \otimes \sqrt{\mathbf{K}_e}\right)\mathbf{R},
\end{align}
which satisfies $\mathbf{L}^\dagger \mathbf{L} = \mathbf{K} / \alpha_K$ where $\alpha_K$ is a constant.
By choosing $\beta=2/\alpha_K$, the shifted stiffness matrix can then be expressed as
\begin{align*}
    \mathbf{B}_F = \mathbf{I} - \beta \mathbf{K} = \mathbf{I} - 2 \mathbf{L}^\dagger \mathbf{L}.
\end{align*}
We exploit this structure to construct a normalized block-encoding by conjugating a reflection about the output subspace of a block-encoding of $\mathbf{L}$.
In the subsequent sections, we construct the block-encodings of the square root of the element stiffness matrix $\sqrt{\mathbf{K}_e}$ and the element restriction operator $\mathbf{R}$ to construct the block-encoding of $\mathbf{L}$, which enables the normalized block-encoding of the shifted global stiffness matrix.

\subsection{Block-encoding of the square root of the element stiffness matrix}
\label{subsec:element-block-encoding}

Here, we focus on the element stiffness matrix for two-dimensional materials in a plane stress mode.
For a square $Q_1$ element, exact integration of $\mathbf{B}^{\top} \mathbf{C} \mathbf{B}$ gives
\begin{equation}
    \mathbf{K}_e = \lambda' \widehat{\mathbf{K}}_e(\nu),
    \label{eq:physical-element-matrix}
\end{equation}
where $\lambda'=E/(1-\nu^2)$ is the scaling factor under the plane stress assumption.
We hereinafter notate $\lambda_\mathrm{ps}$ as $\lambda$ for simplicity.
In the conventional counter-clockwise local-node order
\begin{equation}
    (v_\xi, v_\eta) = (0,0),(1,0),(1,1),(0,1),
    \label{eq:counter-clockwise-node-order}
\end{equation}
for local coordinate $(v_\xi, v_\eta)$ with the two displacement components adjacent at each node, the dimensionless matrix can be written as~\cite{sigmund200199}
\begin{equation}
\widehat{\bm{K}}_{e, \mathrm{ccw}} =
\begin{bmatrix}
 k_1&k_2&k_3&k_4&k_5&k_6&k_7&k_8\\
 k_2&k_1&k_8&k_7&k_6&k_5&k_4&k_3\\
 k_3&k_8&k_1&k_6&k_7&k_4&k_5&k_2\\
 k_4&k_7&k_6&k_1&k_8&k_3&k_2&k_5\\
 k_5&k_6&k_7&k_8&k_1&k_2&k_3&k_4\\
 k_6&k_5&k_4&k_3&k_2&k_1&k_8&k_7\\
 k_7&k_4&k_5&k_2&k_3&k_8&k_1&k_6\\
 k_8&k_3&k_2&k_5&k_4&k_7&k_6&k_1
\end{bmatrix},
\label{eq:element-matrix-ccw}
\end{equation}
where
\begin{align}
 k_1&=\frac12-\frac{\nu}{6}, \quad
k_2=\frac18+\frac{\nu}{8}, \nonumber \\
k_3&=-\frac14-\frac{\nu}{12}, \quad
k_4=-\frac18+\frac{3\nu}{8}, \quad
\nonumber\\
 k_5&=-\frac14+\frac{\nu}{12}, \quad
k_6=-\frac18-\frac{\nu}{8}, \nonumber \\
k_7&=\frac{\nu}{6}, \quad
k_8=\frac18-\frac{3\nu}{8}.
\label{eq:element-k-coefficients}
\end{align}

On the other hand, it is convenient to use the local-node sequence
\begin{equation}
    (v_\xi, v_\eta) = (0,0),(1,0),(0,1),(1,1),
\end{equation}
for the quantum register since it corresponds to the binary order 
\begin{align}
    \ket{v_\eta, v_\xi} = \ket{0 0}, \ket{0 1}, \ket{1 0}, \ket{1 1} = \ket{0}, \ket{1}, \ket{2}, \ket{3}.
\end{align}
Thus, we employ this ordering to define the element stiffness matrix.
We define the corresponding quantum register as $\ket{v_\eta}_\eta \ket{v_\xi}_\xi \ket{v_\mathrm{c}}_c$ where $\ket{v_\mathrm{c}}_c$ represents the displacement component and each register includes one qubit.
For example, $\ket{0}_\eta \ket{1}_\xi \ket{0}_c$ corresponds to the basis state associated with the $x$-component of node $(1, 0)$.
We use subscripts to explicitly indicate the corresponding quantum registers and omit them whenever they are clear from the context.

A central property of the element stiffness matrix is that it is diagonalized by a unitary independent of $\nu$. 
Let
\begin{align}
    \mathbf{U}_{\eta, c} &:= \mathbf{H}_{\eta} \mathbf{Z}_{\eta} C_{c}(\mathbf{X}_{\eta}) C_{\eta}(\mathbf{X}_c) \nonumber \\
    &\times \mathbf{R}_{\mathrm{y}, \eta} \left( -\frac{3}{4} \pi \right) \otimes \mathbf{R}_{\mathrm{y}, c} \left( -\frac{3}{4} \pi \right)
    \label{eq:U4}
\end{align}
acting on the $(\eta, c)$ register where $\mathbf{H}$ is the Hadamard gate, $\mathbf{Z}$ is the Z gate, $\mathbf{R}_\mathrm{y}$ is the Ry gate with the subscript describing the target quantum register, and $C_{a} (\mathbf{X}_{b})$ is the controlled-NOT (CNOT) gate acting on $b$ register conditioned on $a$ register with $a, b \in (\eta, c)$.
Let us define
\begin{equation}
    \mathbf{V}
    := C_{\xi}(\mathbf{U}_{\eta, c}) \,
    C_{\eta} (\mathbf{X}_{\xi}) \,
    \left(\mathbf{H}_{\eta} \otimes \mathbf{H}_{\xi} \otimes \mathbf{H}_c \right),
    \label{eq:element-diagonalizer}
\end{equation}
where
\begin{equation}
    C_{\xi}(\bm{U}_{\eta, c})
    =\ket{0} \! \bra{0}_{\xi} \otimes \mathbf{I}_{\eta, c}
    +\ket{1} \! \bra{1}_{\xi} \otimes \mathbf{U}_{\eta, c},
\end{equation}
with $\mathbf{I}_{\eta, c}$ the identity operator acting on the $(\eta, c)$ register.
Direct multiplication gives the diagonalization of the element stiffness matrix as
\begin{equation}
    \mathbf{D}_e = \mathbf{V} \widehat{\mathbf{K}}_e \mathbf{V}^{\dagger},
    \label{eq:element-diagonalization}
\end{equation}
where 
\begin{equation}
    \mathbf{D}_e =
    \texttt{diag} \!\left(
    0,0,0,1-\nu, k_1, k_1, 1-\nu, 1+\nu
    \right).
    \label{eq:element-diagonal}
\end{equation}
The three zero eigenvalues represent the two rigid translations and one in-plane rigid rotation of an unconstrained element. 

Based on this diagonalization, we now consider the block-encoding of the square root of the element stiffness matrix.
First, we construct the block-encoding of the diagonal matrix $\sqrt{\mathbf{D}}_e$, where we need one ancilla qubit $a_e$. 
The minimum exact normalization for the diagonal matrix is $\sqrt{\alpha_e}$ where $\alpha_e$ is given as
\begin{equation}
    \alpha_e = \| \mathbf{D}_e \|
    =\max\left\{
    \lvert 1-\nu \rvert , \lvert k_1 \rvert,\lvert 1+\nu \rvert
    \right\}.
    \label{eq:alpha-element-general}
\end{equation}
For the usual nonnegative effective Poisson ratio, $0\leq\nu<1$, this reduces to
\begin{equation}
    \alpha_e = 1+\nu.
    \label{eq:alpha-element-physical}
\end{equation}
Let $\ket{r}$ be the collection of quantum registers $\ket{v_\eta} \! \ket{v_\xi} \! \ket{v_\mathrm{c}}$ consisting of three qubits.
For each computational basis state $\ket{r}$, apply
\begin{equation}
    \mathbf{R}_\mathrm{y} \! \left(2\arccos \sqrt{\frac{\omega_r}{\alpha_e}} \right)
    \label{eq:multiplexed-ry-angle}
\end{equation}
for $a_e$, conditioned on $r$, where $\omega_r$ is the $r$-th diagonal entry of Eq.~\eqref{eq:element-diagonal}. 
The resulting multiplexed unitary is given as
\begin{equation}
    \mathbf{U}_{\sqrt{D}} = \sum_{r=0}^{7} \mathbf{R}_\mathrm{y} \! \left(2\arccos \sqrt{\frac{\omega_r}{\alpha_e}} \right) \otimes \ket{r} \! \bra{r},
    \label{eq:UD-element}
\end{equation}
which satisfies
\begin{equation}
    \left(\bra{0}_{a_e} \otimes \mathbf{I}_{\eta, \xi, c} \right)
    \mathbf{U}_{\sqrt{D}}
    \left(\ket{0}_{a_e} \otimes \mathbf{I}_{\eta, \xi, c} \right)
    = \sqrt{\frac{\mathbf{D}_e}{\alpha_e}}.
    \label{eq:UD-top-block}
\end{equation}
Since $\mathbf{V}$ is unitary, we obtain the $(\sqrt{\alpha_e}, 1, 0)$-block-encoding of the square root of the element stiffness matrix $\sqrt{\widehat{\bm{K}}_e}$ as
\begin{equation}
    \mathbf{U}_{\sqrt{e}} =
    \left(\mathbf{I}_{a_e} \otimes \mathbf{V}^{\dagger} \right)
    \mathbf{U}_{\sqrt{D}}
    \left(\mathbf{I}_{a_e} \otimes \mathbf{V} \right),
    \label{eq:element-block-encoding-unitary}
\end{equation}
which satisfies
\begin{equation}
    \left(\bra{0}_{a_e} \otimes \mathbf{I}_{\eta, \xi, c} \right)
    \mathbf{U}_{\sqrt{e}}
    \left(\ket{0}_{a_e} \otimes \mathbf{I}_{\eta, \xi, c} \right)
    = \sqrt{\frac{\widehat{\bm{K}}_e}{\alpha_e}}.
    \label{eq:element-block-encoding-property}
\end{equation}
The physical matrix $\mathbf{K}_e = \lambda \widehat{\mathbf{K}}_e$ is therefore represented with normalization $\sqrt{\lambda \alpha_e}$. 
The gate count for implementing $\mathbf{U}_{\sqrt{e}}$ is provided in Appendix~\ref{sec:count_sqrt_Ke}.

Although this study focuses on the explicit quantum circuit construction for two-dimensional problems, the same approach can be extended to three dimensions.
For a trilinear hexahedral element of side length $h$, the element stiffness matrix scales as
$\mathbf{K}_e^{(3)} = Eh\widehat{\mathbf{K}}_e^{(3)}$,
where $\widehat{\mathbf{K}}_e^{(3)}$ is a dimensionless $24\times24$ element stiffness matrix.
Consequently, the corresponding block-encoding normalization
for the square root of the element stiffness matrix
scales as $O(\sqrt{h})$ for fixed material properties.

\subsection{Block-encoding of the element restriction operator}
\label{subsubsec:R-block-encoding}

\begin{figure}[t]
	\centering
	\includegraphics[width=\columnwidth]{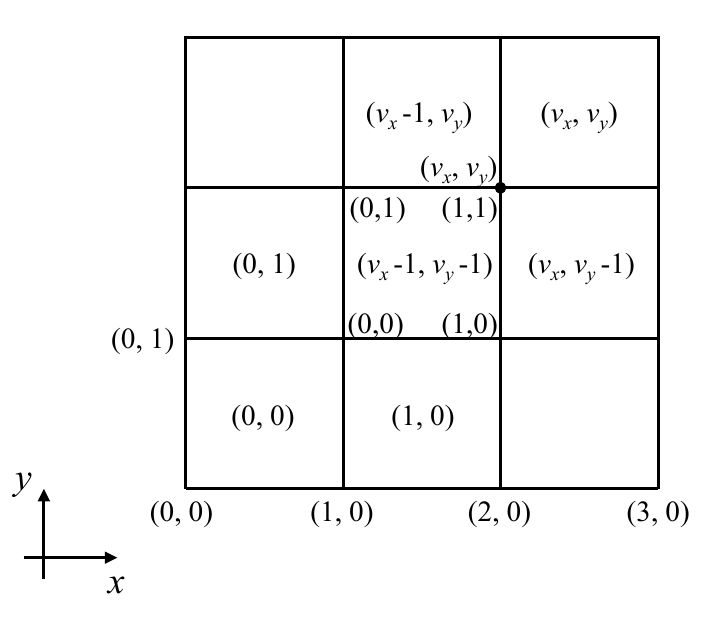}
	\caption{Relationship between node and element labels. The values at the nodes represent the node labels $(v_x, v_y)$, while the values inside the elements represent the element labels $(e_x, e_y)$. The four numbers at the corners of the central element represent the local node labels $(v_\xi, v_\eta)$ within the element $(e_x, e_y) = (v_x-1, v_y-1)$. The node $(v_x, v_y)$ belongs to the four surrounding elements. For the lower-left element, $(e_x, e_y) = (v_x-1, v_y-1)$, the corresponding local node is $(v_\xi, v_\eta) = (1,1)$.} \label{fig:vertex-element}
\end{figure}

Here, we construct the block-encoding of the element restriction operator $\mathbf{R}$.
As shown in Fig.~\ref{fig:vertex-element}, the node $(v_\mathrm{x}, v_\mathrm{y})$ belongs to four possible neighboring elements at most with the labels of
\begin{equation}
    e_\mathrm{x} = v_\mathrm{x} - v_\xi,
    \qquad
    e_\mathrm{y} = v_\mathrm{y} - v_\eta,
    \label{eq:node-to-element-coordinates}
\end{equation}
where $(v_\xi, v_\eta) \in \{0, 1\}^2$. 
The neighboring element exists if
\begin{equation}
    0 \leq v_\mathrm{x} - v_\xi < N_\mathrm{x},
    \qquad
    0\leq v_\mathrm{y} - v_\eta < N_\mathrm{y}.
    \label{eq:occurrence-validity}
\end{equation}
In the following, we exploit this regularity of the structured mesh to derive an explicit representation of the element restriction operator, i.e., the global-to-local mapping.

Let $b(v_\xi, v_\eta, v_\mathrm{x}, v_\mathrm{y}) \in \{0,1\}$ equal zero for a valid case satisfying Eq.~\eqref{eq:occurrence-validity} and one for an invalid case.
The quantum circuit construction for the element restriction operator uses the quantum registers denoted by
\begin{equation}
    (w, b, l, v), \qquad l = (\xi, \eta), \qquad v = (x, y, c),
    \label{eq:R-registers}
\end{equation}
where $b$ has one qubit for the invalid case flag, $w$ consists of two work qubits, $x$ and $y$ are $\lceil \log_2 (N_\mathrm{x} + 1) \rceil$ and $\lceil \log_2 (N_\mathrm{y} + 1) \rceil$ qubit registers, respectively. 
Starting from $\ket{v_\eta, v_\xi}=\ket{00}$, Hadamard gates prepare the uniform superposition of four neighboring element labels. 
Reversible equality tests compute lower- and upper-bound violations, and controlled decrementers implement Eq.~\eqref{eq:node-to-element-coordinates}. 
The action is written as
\begin{align}
    &\ket{0}_{b} \ket{00}_l \ket{v_\mathrm{y}, v_\mathrm{x}, c}_v
    \nonumber\\
    &\xrightarrow{\mathbf{H}^{\otimes 2}_{\xi, \eta}} \frac{1}{2} \sum_{v_\xi, v_\eta \in \{0,1\}}
    \ket{0}_{b} \ket{v_\eta, v_\xi}_l \ket{v_\mathrm{y}, v_\mathrm{x}, c}_v \nonumber \\
    &\xrightarrow{\mathbf{U}_\mathrm{flag}} \frac{1}{2} \sum_{v_\xi, v_\eta \in \{0,1\}}
    \ket{b}_{b} \ket{v_\eta, v_\xi}_l \ket{v_\mathrm{y}, v_\mathrm{x}, v_c}_v \nonumber \\
    &\xrightarrow{\mathbf{U}_\mathrm{decr}} \frac{1}{2} \sum_{v_\xi, v_\eta \in \{0,1\}}
    \ket{b}_{b} \ket{v_\eta, v_\xi}_l \nonumber \\
    & \quad \quad \quad \quad \quad \quad \quad \quad \quad \quad \otimes \ket{v_\mathrm{y} - v_\eta, v_\mathrm{x} - v_\xi, v_c}_v \nonumber \\
    &= \frac{1}{2} \sum_{v_\xi, v_\eta \in \{0,1\}}
    \ket{b}_{b} \ket{v_\eta, v_\xi}_l \ket{e_\mathrm{y}, e_\mathrm{x}, v_c}_v,
    \label{eq:US-action}
\end{align}
where we use the working qubits $w$ to implement $\mathbf{U}_\mathrm{flag}$.
Specifically, $\mathbf{U}_\mathrm{flag}$ is implemented as follows.
Let $\mathbf{U}_{x, y ,l \rightarrow w}$ denote 
\begin{align}
    \mathbf{U}_{x, y, l \rightarrow w} &:= C_{v_\mathrm{y}=N_\mathrm{y}, v_\eta=0} (\mathbf{X}_{w_1}) C_{v_\mathrm{y}=0, v_\eta=1} (\mathbf{X}_{w_1}) \nonumber \\
    & \times C_{v_\mathrm{x}=N_\mathrm{x}, v_\xi=0} (\mathbf{X}_{w_0}) C_{v_\mathrm{x}=0, v_\xi=1} (\mathbf{X}_{w_0}),
\end{align}
where $w=(w_0, w_1)$, and $C_{v_\mathrm{x}=0, v_\xi=1} (\mathbf{X}_{w_0})$ represents the multi-controlled $\mathbf{X}$ gate acting on the qubit $w_0$ conditioned on the $x$ and $\xi$ registers being in the states $\ket{v_\mathrm{x}} = \ket{0}$ and $\ket{v_\xi} = \ket{1}$, respectively.
The other controlled gates with different subscripts are interpreted analogously, that is, the subscript specifies the control register and the value of the corresponding register on which the operation is conditioned.
By this gate, the state of the qubit $w_0$ is flipped if $\ket{v_\mathrm{x}} = \ket{0}$ and $\ket{v_\xi} = \ket{1}$ or $\ket{v_\mathrm{x}} = \ket{N_\mathrm{x}}$ and $\ket{v_\xi} = \ket{0}$, and the state of the qubit $w_1$ is flipped if $\ket{v_\mathrm{y}} = \ket{0}$ and $\ket{v_\eta} = \ket{1}$ or $\ket{v_\mathrm{y}} = \ket{N_\mathrm{y}}$ and $\ket{v_\eta} = \ket{0}$.
These conditions correspond to the invalid branches where the conditions in Eq.~\eqref{eq:occurrence-validity} do not hold.
Hence, the invalid flag is set for $b = w_0 \lor w_1$, which can be implemented as
\begin{align}
    \mathbf{U}_{w \rightarrow b} := C_{w_0, w_1} (\mathbf{X}_{b}) C_{w_1} (\mathbf{X}_{b}) C_{w_0} (\mathbf{X}_{b}),
\end{align}
where $C_{w_0, w_1} (\mathbf{X}_{b})$ is the Toffoli gate acting on the qubit $b$ conditioned on the qubits $w_0$ and $w_1$.
Together with the uncomputation for the $w$ register, the operator $\mathbf{U}_\mathrm{flag}$ is implemented as
\begin{align}
    \mathbf{U}_\mathrm{flag} = \mathbf{U}_{x, y, l \rightarrow w}^\dagger \mathbf{U}_{w \rightarrow b} \mathbf{U}_{x, y, l \rightarrow w}.
\end{align}
A controlled decrement operator $\mathbf{U}_\mathrm{decr}$ acts on the $x$ and $y$ registers conditioned on the values $v_\xi$ and $v_\eta$, respectively.
A naive $k$-qubit decrementer can be implemented as a ladder of multi-controlled Toffoli gates. 
After the least significant bit is flipped, the $j$-th bit is flipped only when all $j$ less significant bits are in the state $\ket{1}$, thereby correctly propagating the borrow. 
This construction is equivalent to the inverse of the standard multi-controlled Toffoli ladder used for an $n$-qubit incrementer~\cite{khattar2025rise}.
A single-qubit controlled $k$-qubit decrementer can be obtained from a $(k+1)$-qubit decrementer by treating the control qubit as the least significant bit of the extended register. 
An $\mathbf{X}$ gate is applied to the control qubit immediately before the decrementer, canceling the initial bit flip in the decrementer circuit. 
The remaining multi-controlled Toffoli ladder then propagates the borrow only when the control qubit is in the state $\ket{1}$, thereby implementing $\ket{a} \ket{j} \mapsto \ket{a} \ket{(j-a)\bmod 2^k}$.
While this naive implementation of the $k$-qubit decrementer requires $\mathcal{O}(k^2)$ Toffoli gates, Khattar and Gidney provided a more sophisticated implementation where a $k$-qubit incrementer can be implemented using $3k$ Toffoli gates and $\log_{2}^{*}k$ clean ancilla qubits~\cite{khattar2025rise} with $\log_{2}^{*}k$ denoting the iterated logarithm. 
Since a decrementer is the inverse of an incrementer, a controlled $k$-qubit decrementer can be implemented by $3(k+1)$ Toffoli gates.

Let $\mathbf{\Pi}_{\mathrm{in}}$ select $\ket{v_\eta, v_\xi} = \ket{00}$, $\ket{b}=\ket{0}$, and $\ket{w}=\ket{00}$ at the input, and let $\mathbf{\Pi}_{\mathrm{out}}$ select $\ket{b}=\ket{0}$ and $\ket{w}=\ket{00}$ at the output while retaining $\ket{v_\eta, v_\xi}$ as part of the element-local index. 
Let $\mathbf{U}_R$ denote the action of Eq.~\eqref{eq:US-action}, which implies
\begin{equation}
    \mathbf{\Pi}_{\mathrm{out}} \mathbf{U}_R \mathbf{\Pi}_{\mathrm{in}}
    = \frac{\mathbf{R}}{2}.
    \label{eq:R-block-encoding}
\end{equation}
Since $\| \mathbf{R}^{\top} \mathbf{R} \| = 4$ holds, $\| \mathbf{R} \| = 2$, which implies that the above block-encoding has the optimal normalization constant. 
The gate count for implementing $\mathbf{U}_R$ is given in Appendix~\ref{sec:count_restriction}.

\subsection{Dirichlet boundary condition}
\label{subsec:dirichlet}

The global stiffness matrix constructed in Eq.~\eqref{eq:assembled-homogeneous} is unconstrained, that is, the Neumann boundary conditions are imposed on all boundaries.
Here, we discuss the implementation of the Dirichlet boundary conditions on boundaries $\Gamma_\mathrm{D}$.

Let $\mathcal{D}$ be the set of constrained degrees of freedom.
Based on this set, we define the following projectors:
\begin{equation}
    \mathbf{\Pi}_\mathrm{D} = \sum_{q \in \mathcal{D}} \ket{q} \! \bra{q}_v,
    \qquad
    \mathbf{\Pi}_\mathrm{F} = \mathbf{I}-\mathbf{\Pi}_\mathrm{D}.
    \label{eq:PD-PF}
\end{equation}
For example, if the left edges are supposed to be fixed, $\mathcal{D}$ is set as
\begin{equation}
    \mathcal{D} = \{(0, v_\mathrm{y}, v_\mathrm{c}) ~|~ 0 \leq v_\mathrm{y} \leq N_\mathrm{y}, v_\mathrm{c} \in \{0,1\} \}.
    \label{eq:left-dirichlet-set}
\end{equation}
To impose the homogeneous Dirichlet boundary condition $\mathbf{u}=0$ on $\Gamma_\mathrm{D}$, it suffices to replace the global stiffness matrix with
\begin{align}
    \mathbf{K}_\mathrm{F} := \mathbf{\Pi}_\mathrm{F} \mathbf{K} \mathbf{\Pi}_\mathrm{F}.
\end{align}
To this end, we construct the block-encoding of the projector $\mathbf{\Pi}_\mathrm{F}$.

Let us consider a membership oracle that acts on node register $v$ and one ancilla qubit $a_\mathrm{D}$ as
\begin{equation}
    \mathbf{O}_\mathrm{D} \ket{q}_v \ket{z}_{a_\mathrm{D}}
    = \ket{q}_v \ket{z \oplus \chi_\mathrm{D}(q)}_{a_\mathrm{D}},
    \label{eq:membership-oracle}
\end{equation}
for $z \in \{0, 1 \} $ where $\chi_\mathrm{D}(q) = 1$ if $q \in \mathcal{D}$ and zero otherwise. 
Projecting the ancilla qubit state to $\ket{0}_{a_\mathrm{D}}$ gives
\begin{equation}
    \left(\bra{0}_{a_\mathrm{D}} \otimes \mathbf{I}_v \right)
    \mathbf{O}_\mathrm{D}
    \left(\ket{0}_{a_\mathrm{D}} \otimes \mathbf{I}_v \right)
    = \mathbf{\Pi}_\mathrm{F},
    \label{eq:PF-from-membership}
\end{equation}
which means that $\mathbf{O}_\mathrm{D}$ is exactly the $(1, 1, 0)$-block-encoding of $\mathbf{\Pi}_\mathrm{F}$.
For the case of Eq.~\eqref{eq:left-dirichlet-set}, $\chi_\mathrm{D}(q) =1$ if $v_\mathrm{x} = 0$, and thus the oracle is a multi-controlled NOT gate acting on $a_\mathrm{D}$ qubit conditioned on the zero state of the $v_\mathrm{x}$ register.

For nonhomogeneous prescribed displacement $\mathbf{g}$, one may use a lifting technique. 
By introducing a vector $\mathbf{w}$ which satisfies
\begin{equation}
    \mathbf{w} = \mathbf{u} - \mathbf{g},
    \qquad
    \mathbf{\Pi}_\mathrm{D} \mathbf{w} = 0,
\end{equation}
we formulate the systems of linear equations for the vector $\mathbf{w}$, which satisfies the homogeneous Dirichlet boundary condition, as follows.
\begin{equation}
    \mathbf{K}_\mathrm{F} \mathbf{w} = \mathbf{\Pi}_\mathrm{F} (\mathbf{f} - \mathbf{K} \mathbf{g})
    \label{eq:nonhomogeneous-lifting}
\end{equation}
While this requires modified state preparation for the right-hand side, we can use the same block-encoding technique of the stiffness matrix in the left-hand side.

\subsection{Block encoding of a Gram factor}
\label{subsubsec:gram-factor-encoding}

Using Eqs.~\eqref{eq:R-block-encoding} and \eqref{eq:PD-PF}, we define the rectangular matrix
\begin{equation}
    \mathbf{L}
    :=
    \left( \mathbf{I}^{(N_\mathrm{el})} \otimes \sqrt{\frac{\widehat{\mathbf{K}}_e}{\alpha_e}} \right)
    \frac{\mathbf{R}}{2}
    \mathbf{\Pi}_{\mathrm{F}}.
    \label{eq:gram-factor-L}
\end{equation}
Because $\|\mathbf{R}/2\|=1$ and $\|\mathbf{\Pi}_{\mathrm{F}}\|=1$, the matrix $\mathbf{L}$ is a contraction.
Moreover, it holds that
\begin{align}
    \mathbf{L}^{\dagger}\mathbf{L}
    &=
    \mathbf{\Pi}_{\mathrm{F}}
    \frac{\mathbf{R}^{\top}}{2}
    \left(
    \frac{\mathbf{I}^{(N_\mathrm{el})} \otimes \widehat{\mathbf{K}}_e}{\alpha_e}
    \right)
    \frac{\mathbf{R}}{2}
    \mathbf{\Pi}_{\mathrm{F}}
    \nonumber\\
    &=
    \frac{\mathbf{K}_\mathrm{F}}
    {4\lambda (1 + \nu)}.
    \label{eq:gram-factor-property}
\end{align}

Let $\mathbf{U}_L$ be obtained by applying the membership oracle $\mathbf{O}_{\mathrm{D}}$, the block-encoding of the restriction operator, $\mathbf{U}_R$, and the block-encoding of the square root of the element stiffness matrix, $\mathbf{U}_{\sqrt{e}}$, controlled by the zero state of $b$, which corresponds to the valid branch.
At the input, the good ancilla subspace is specified by
\begin{equation}
    (a_\mathrm{D},l,b,w,a_e)
    =(0,00,0,00,0),
\end{equation}
whereas at the output it is specified by
\begin{equation}
    (a_\mathrm{D},b,w,a_e)
    =(0,0,00,0).
\end{equation}
The local register $l=(\xi,\eta)$ is included as part of the output index.
Using the corresponding projectors, respectively denoted by $\mathbf{\Pi}_{L,\mathrm{in}}$ and $\mathbf{\Pi}_{L,\mathrm{out}}$, we obtain the block-encoding
\begin{equation}
    \mathbf{\Pi}_{L,\mathrm{out}}
    \mathbf{U}_L
    \mathbf{\Pi}_{L,\mathrm{in}}
    =\mathbf{L}.
    \label{eq:L-block-encoding}
\end{equation}
This construction requires five ancilla qubits: one membership flag $a_\mathrm{D}$, one invalid flag $b$, two work qubits $w$, and one square-root block-encoding qubit $a_e$.
The gate count for implementing $\mathbf{U}_L$ is discussed in Appendix~\ref{sec:count_gram}.

For the heterogeneous case when the Young's modulus is element-wise constant, we replace the above $\mathbf{I}^{(N_\mathrm{el})}$ in Eq.~\eqref{eq:gram-factor-L} with $\mathbf{D}_{\sqrt{\rho}}$, which is a diagonal matrix whose entries encode the Young's modulus at each element, assuming the block-encoding of $\mathbf{D}_{\sqrt{\rho}}$.
A more detailed discussion is provided in Appendix~\ref{sec:heterogeneous-material}.

\subsection{Reflection operator for a normalized block-encoding}
\label{subsubsec:normalized-B-reflection}

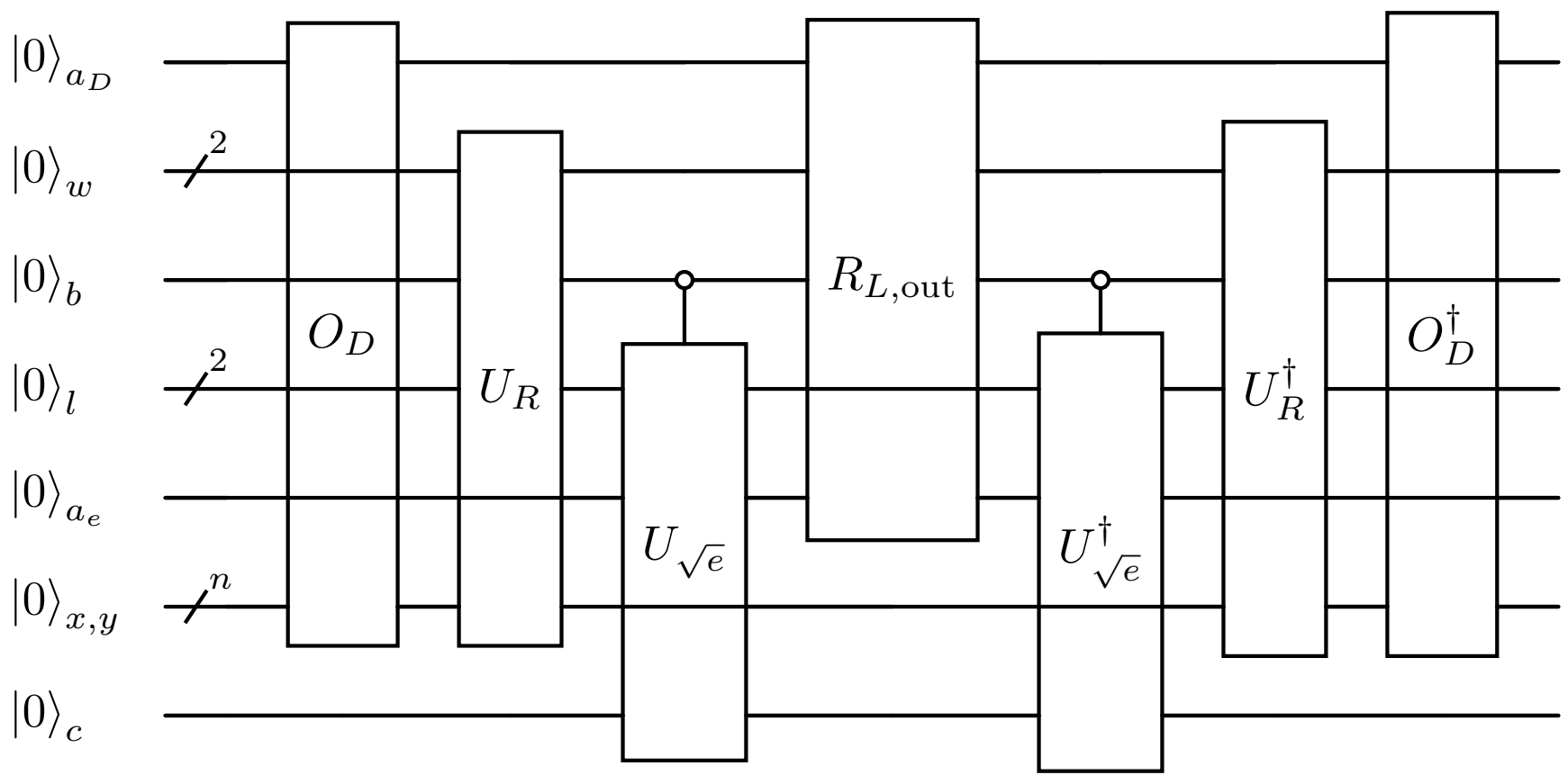
\begin{figure*}[t]
    \centering
    \begin{quantikz}[column sep=0.45cm, row sep={0.8cm,between origins}, transparent]
        \lstick{$\makebox[1cm][l]{$\ket{0}_{a_D}$}$} & \qw & \gate[6]{O_D} &  & \qw & \gate[5]{R_{L,\mathrm{out}}} & \qw & \qw & \gate[6]{O_D^\dagger} & \qw \\
        \lstick{$\makebox[1cm][l]{$\ket{0}_{w}$}$} & \qwbundle{2} & \linethrough & \gate[5]{U_R} & \qw & \qw & \qw & \gate[5]{U_R^\dagger} & \linethrough & \qw \\
        \lstick{$\makebox[1cm][l]{$\ket{0}_{b}$}$} & \qw & \linethrough & \qw & \octrl{1} & \qw & \octrl{1} & \qw & \linethrough & \qw \\
        \lstick{$\makebox[1cm][l]{$\ket{0}_{l}$}$} & \qwbundle{2} & \linethrough & \qw & \gate[4]{U_{\sqrt e}} & \linethrough & \gate[4]{U_{\sqrt e}^\dagger} & \qw & \linethrough & \qw \\
        \lstick{$\makebox[1cm][l]{$\ket{0}_{a_e}$}$} & \qw & \linethrough & \linethrough & \qw & \qw & \qw & \linethrough & \linethrough & \qw \\
        \lstick{$\makebox[1cm][l]{$\ket{0}_{x, y}$}$} & \qwbundle{n} & \qw & \qw & \linethrough & \qw & \linethrough & \qw & \qw & \qw \\
        \lstick{$\makebox[1cm][l]{$\ket{0}_{c}$}$} & \qw & \qw & \qw & \qw & \qw & \qw & \qw & \qw & \qw \\
    \end{quantikz}
    \caption{Quantum circuit for the normalized block encoding $U_B = U_L^\dagger R_{L,\mathrm{out}} U_L$. Here, $n := n_x + n_y$.}
    \label{fig:block-encoding-B}
\end{figure*}

Let $\mathbf{R}_{L,\mathrm{out}}$ denote the reflection about the output-good subspace defined as
\begin{equation}
    \mathbf{R}_{L,\mathrm{out}}
    :=\mathbf{I}-2\mathbf{\Pi}_{L,\mathrm{out}}.
    \label{eq:L-output-reflection}
\end{equation}
The unitary
\begin{equation}
    \mathbf{U}_{B}
    :=
    \mathbf{U}_{L}^{\dagger}
    \mathbf{R}_{L,\mathrm{out}}
    \mathbf{U}_{L}
    \label{eq:normalized-B-unitary}
\end{equation}
then satisfies
\begin{align}
    &\mathbf{\Pi}_{L,\mathrm{in}}
    \mathbf{U}_{B}
    \mathbf{\Pi}_{L,\mathrm{in}} \nonumber \\
    &=
    \mathbf{\Pi}_{L,\mathrm{in}}
    \mathbf{U}_{L}^{\dagger}
    (\mathbf{I}-2\mathbf{\Pi}_{L,\mathrm{out}})
    \mathbf{U}_{L}
    \mathbf{\Pi}_{L,\mathrm{in}} \nonumber \\
    &=
    \mathbf{I}
    -2
    \left(
    \mathbf{\Pi}_{L,\mathrm{out}}
    \mathbf{U}_{L}
    \mathbf{\Pi}_{L,\mathrm{in}}
    \right)^{\dagger}
    \left(
    \mathbf{\Pi}_{L,\mathrm{out}}
    \mathbf{U}_{L}
    \mathbf{\Pi}_{L,\mathrm{in}}
    \right)
    \nonumber\\
    &=
    \mathbf{I}-2\mathbf{L}^{\dagger}\mathbf{L}
    \nonumber\\
    &=
    \mathbf{I}
    -\frac{
    \mathbf{K}_{\mathrm{F}}
    }{
    2\lambda ( 1 + \nu)
    },
    \label{eq:normalized-B-property}
\end{align}
where we use $\mathbf{\Pi}_{L,\mathrm{out}} = \mathbf{\Pi}_{L,\mathrm{out}}^2$, $\mathbf{\Pi}_{L,\mathrm{out}} = \mathbf{\Pi}_{L,\mathrm{out}}^\dagger$, and $\mathbf{\Pi}_{L,\mathrm{in}} = \mathbf{\Pi}_{L,\mathrm{in}}^\dagger$.
Consequently, $\mathbf{U}_B$ is a normalized block-encoding of
\begin{equation}
    \mathbf{B}_{\mathrm{F}}
    =\mathbf{I}-\beta \mathbf{K}_{\mathrm{F}},
    \label{eq:eta-FEM}
\end{equation}
where 
\begin{align}
    \beta = \frac{1}{2\lambda (1 + \nu)}
\end{align}
The unitary $\mathbf{U}_{B}$ uses the seven ancilla qubits: one membership flag $a_\mathrm{D}$, one invalid flag $b$, two work qubits $w$, two local qubits $l$, and one square-root block-encoding qubit $a_e$.
Figure~\ref{fig:block-encoding-B} shows the quantum circuit for block-encoding of $\mathbf{B}_\mathrm{F}$.
The gate count for implementing the normalized shifted stiffness matrix is provided in Appendix~\ref{sec:count_reflection}.
The normalized block-encoding of the shifted stiffness matrix has the gate complexity of $\mathcal{O}(\log N)$ up to the cost of the material and membership oracles.

\subsection{Complexity analysis}
\label{subsec:solving-OD}

\subsubsection{Displacement state preparation}

The matrix $\mathbf{K}_\mathrm{F}$ is singular on the full global space because every constrained basis state belongs to its kernel.
On the free subspace, however, sufficient Dirichlet constraints make it positive definite.
For a load satisfying
\begin{equation}
    \mathbf{\Pi}_\mathrm{F}\mathbf{f}=\mathbf{f},
    \label{eq:free-force}
\end{equation}
the physical solution is
\begin{equation}
    \mathbf{u}=\mathbf{K}_\mathrm{F}^{+}\mathbf{f},
    \qquad
    \mathbf{\Pi}_\mathrm{D}\mathbf{u}=0.
    \label{eq:dirichlet-pseudoinverse-solution}
\end{equation}

We now derive the complexity to prepare the state encoding the displacement.
Let $\varsigma_\mathrm{min}^{+}(\mathbf{K}_\mathrm{F})$ be the smallest nonzero eigenvalue of $\mathbf{K}_\mathrm{F}$ and $\kappa_B^\star$ be set as
\begin{equation}
    \kappa_B
    \geq
    \kappa_B^\star
    :=
    \frac{1}{
    \beta \varsigma_\mathrm{min}^{+}(\mathbf{K}_\mathrm{F})
    }
    =
    \frac{
    2\lambda(1+\nu)
    }{
    \varsigma_\mathrm{min}^{+}(\mathbf{K}_\mathrm{F})
    }
    \label{eq:kappa-B-FEM}
\end{equation}
for the homogeneous case.
Since
\begin{equation}
    \varsigma_\mathrm{max}(\mathbf{K}_{\mathrm F})
    \leq 4\lambda(1+\nu), \label{eq:max-eigenvalue}
\end{equation}
the effective condition number satisfies
\begin{equation}
    \kappa(\mathbf{K}_{\mathrm F})
    :=
    \frac{\varsigma_\mathrm{max}(\mathbf{K}_{\mathrm F})}
    {\varsigma_\mathrm{min}^{+}(\mathbf{K}_{\mathrm F})}
    \leq 2\kappa_B^\star.
    \label{eq:kappa_upper_bound}
\end{equation}
On the other hand, the maximum eigenvalue of the stiffness matrix is lower bounded by its diagonal entry, that is, 
\begin{align}
    \varsigma_{\max}(\mathbf{K}_F)
    \geq
    2\lambda\left(1-\frac{\nu}{3}\right).
\end{align}
Thus, we obtain
\begin{equation}
    \kappa(\mathbf{K}_{\mathrm F})
    =
    \frac{\varsigma_\mathrm{max}(\mathbf{K}_{\mathrm F})}
    {\varsigma_\mathrm{min}^{+}(\mathbf{K}_{\mathrm F})}
    \geq \frac{1-\frac{\nu}{3}}
    {1 + \nu} \kappa_B^\star.
    \label{eq:kappa-lower-bound}
\end{equation}
Consequently, $\kappa_B^\star \in \Theta(\kappa)$.

Applying the polynomial in Eq.~\eqref{eq:OD-inverse-polynomial} to the normalized block-encoding $\mathbf{U}_B$ approximates the inverse on the free subspace and maps the constrained subspace to zero.
Let $\mathbf{U}_{K_\mathrm{F}^{+}}$ denote the resulting unitary.
With
\begin{equation}
    \alpha_{K_\mathrm{F}^{+}}
    :=\beta P_\mathrm{max},
    \label{eq:OD-pseudoinverse-normalization}
\end{equation}
it satisfies
\begin{equation}
    \left\|
    \mathbf{K}_\mathrm{F}^{+}
    -
    \alpha_{K_\mathrm{F}^{+}}
    \left(\bra{0^{a_+}}\otimes\mathbf{I}\right)
    \mathbf{U}_{K_\mathrm{F}^{+}}
    \left(\ket{0^{a_+}}\otimes\mathbf{I}\right)
    \right\|
    \leq \beta \epsilon_\mathrm{poly},
    \label{eq:Kplus-block-encoding}
\end{equation}
where $\epsilon_\mathrm{poly}$ is the polynomial approximation error.
The implementation has seven ancilla qubits for $\mathbf{U}_B$, one for QSVT, one for linear combination of unitaries combining the even and odd polynomials, and thus, $a_+=9$.
To bound the error of the block-encoding $\mathbf{U}_{K_\mathrm{F}^+}$ within $\epsilon$, it suffices to set $\epsilon_\mathrm{poly} \leq \epsilon / \beta$.
Consequently, the number of calls to $\mathbf{U}_B$ or $\mathbf{U}_B^{\dagger}$ for solving the system of linear equations reads
\begin{equation}
    \mathcal{O}\!\left(
    \sqrt{\kappa}
    \log\frac{\kappa}{2\lambda (1+\nu) \epsilon}
    \right),
    \label{eq:OD-FEM-query-complexity}
\end{equation}
where we use $\beta = 1/(2 \lambda (1 + \nu))$ and $\kappa \in \Theta(\kappa_B^\star)$.
By applying $\mathbf{U}_{K_\mathrm{F}^{+}}$ to the amplitude-encoded force state $\ket{f}=\mathbf{f}/\|\mathbf{f}\|$, we obtain
\begin{align}
    \mathbf{U}_{K_\mathrm{F}^{+}}
    \ket{0^{a_+}}\ket{f}
    & =
    \ket{0^{a_+}}
    \frac{
    \mathbf{u} + \mathbf{E}\mathbf{f}
    }{
    \alpha_{K_\mathrm{F}^{+}}\|\mathbf{f}\|
    }
    +\ket{\perp},
    \label{eq:solution-state-before-postselection}
\end{align}
where $\|\mathbf{E}\|\leq\epsilon$ and $\ket{\perp}$ is orthogonal to the all-zero ancilla subspace.
Ignoring the small error term, the naive success probability of the postselection is given as
\begin{equation}
    p_{\mathrm{succ}}
    \simeq
    \frac{
    \| \mathbf{u} \|^2
    }{
    \alpha_{K_\mathrm{F}^{+}}^2\|\mathbf{f}\|^2
    },
    \label{eq:OD-postselection-probability}
\end{equation}
which is amplified to $\mathcal{O}(1)$ by amplitude amplification technique with the number of calls of
\begin{equation}
    \mathcal{O}\!\left(
    \frac{
    \alpha_{K_\mathrm{F}^{+}}\|\mathbf{f}\|
    }{
    \|\mathbf{u}\|
    }
    \right) = \mathcal{O}\!\left(
    \frac{
    \kappa \|\mathbf{f}\|
    }{
    \|\mathbf{u}\|
    }
    \right),
    \label{eq:OD-amplitude-amplification-overhead}
\end{equation}
since $\alpha_{K_\mathrm{F}^{+}}=\beta P_\mathrm{max} \in \Theta (\kappa)$.
That is, the square-root dependence in Eq.~\eqref{eq:OD-FEM-query-complexity} is the query complexity for constructing the block-encoding of the pseudoinverse $\mathbf{K}_\mathrm{F}^{+}$.
The end-to-end complexity for preparing the normalized displacement state additionally depends on the norm $\| \mathbf{u} \|$ and $\| \mathbf{f} \|$.

To determine this overhead, consider a fixed physical domain discretized by a quasi-uniform two-dimensional mesh with $N$ degrees of freedom and mesh size $h$, such that
\begin{align}
    N\in\Theta(h^{-2}),
    \qquad
    \kappa\in\Theta(h^{-2})=\Theta(N).
\end{align}
For a mesh-independent displacement field, the nodal coefficient vector satisfies
\begin{align}
    \| \mathbf{u} \| \in \Theta(\sqrt{N}).
\end{align}
In contrast, the scaling of the assembled force vector must include the corresponding finite element integration weights. 
For a fixed body force, each nonzero entry is of order $h^2$, and hence
\begin{align}
    \| \mathbf{f} \|
    \in\Theta(\sqrt{N}h^2)
    =\Theta(N^{-1/2}).
\end{align}
For a fixed boundary traction, each nonzero entry is of order $h$ and the load acts on $\Theta(N^{1/2})$ boundary degrees of freedom, giving
\begin{align}
    \|\mathbf{f}\|
    \in\Theta(N^{1/4} h)
    =\Theta(N^{-1/4}).
\end{align}
For a fixed nodal point force,
\begin{align}
    \|\mathbf{f} \|\in\Theta(1).
\end{align}
Therefore, the success probability scales as
\begin{equation}
    p_{\mathrm{succ}} \in 
        \begin{cases}
            \Theta(1), & \text{body force}, \\ \Theta(N^{-\frac{1}{2}})
            & \text{boundary traction} \\
            \Theta(N^{-1}) 
            & \text{point force},
        \end{cases}
    \end{equation}
and the corresponding overhead reads
\begin{equation}
    \mathcal{O}\left(p_{\mathrm{succ}}^{-1/2}\right) = 
    \begin{cases}
        \mathcal{O}(1),
        & \text{body force} \\
        \mathcal{O}(N^{\frac{1}{4}})
        & \text{boundary traction} \\
        \mathcal{O}(N^{\frac{1}{2}})
        & \text{point force}.
    \end{cases}
\label{eq:OD-load-dependent-AA-overhead}
\end{equation}
Thus, for fixed physical load magnitudes, distributed body forces give the largest success probability of postselection, whereas increasingly localized loads require a larger overhead.

\subsubsection{Strain energy estimation}
\label{subsubsec:strain-energy}

A quantum linear system solver prepares the solution vector $\mathbf{u}$ as amplitude encoding of a quantum state. 
This makes the access to all components of $\mathbf{u}$ as a classical vector prohibitive because of the need for quantum state tomography. 
Therefore, an appropriate strategy is required for extracting meaningful information from the quantum state, based on the physical quantity of interest. 
This subsection gives procedures for estimating strain energy as a global energetic quantity.

For a displacement satisfying the homogeneous Dirichlet condition, the total strain energy $\mathcal{W}$ is given as
\begin{align}
    \mathcal{W}
    &=\frac{1}{2} \mathbf{u}^{\top} \mathbf{K}_\mathrm{F} \mathbf{u} \nonumber \\
    &= \frac{1}{2} \mathbf{f}^{\top}  (\mathbf{K}_\mathrm{F}^{+})^{\top} \mathbf{K}_\mathrm{F} \mathbf{K}_\mathrm{F}^{+} \mathbf{f} \nonumber \\
    &= \frac{1}{2} \mathbf{f}^{\top} \mathbf{K}_\mathrm{F}^{+} \mathbf{f} \nonumber \\
    &= \frac{1}{2} \| \mathbf{f} \|^2 \frac{P_\mathrm{max}}{2\lambda (1 + \nu)} \bra{0^{a_{+}}} \! \bra{f} \mathbf{U}_{K_\mathrm{F}^{+}} \ket{0^{a_{+}}} \! \ket{f}.
    \label{eq:strain-energy}
\end{align}
In this case, we need to estimate the expected value $\bra{0^{a_{+}}} \! \bra{f} \mathbf{U}_{K_\mathrm{F}^{+}} \ket{0^{a_{+}}} \! \ket{f}$.
To estimate the strain energy with accuracy $\epsilon_\mathrm{exp}$, we need to estimate the expected value within error
\begin{align}
    \frac{4 \lambda(1+\nu)}{\| \mathbf{f} \|^2 P_\mathrm{max}} \epsilon_\mathrm{exp},
\end{align}
which requires the amplitude estimation with
\begin{align}
    \mathcal{O} \left( \frac{\| \mathbf{f} \|^2 P_\mathrm{max}}{\lambda(1+\nu) \epsilon_\mathrm{exp}} \right)
\end{align}
calls to $\mathbf{U}_{K_\mathrm{F}^{+}}$.

\subsubsection{Low-resolution tomography}
\label{subsubsec:low-resolution-tomography}

Full tomography of the displacement state $\ket{u} = \mathbf{u} / \| \mathbf{u} \|$ would require a number of measurements that scale with the number of grids and would cancel out the potential quantum advantage of quantum linear system solvers. 
Instead, we herein consider reconstructing only a prescribed coarse displacement field. 
Let the fine node set be partitioned into $M_\mathrm{x} \times M_\mathrm{y}$ disjoint rectangular blocks and $\mathcal{B}_{pq}$ denote the block at the $p$-th position in the $x$-direction and the $q$-th position in the $y$-direction, where both indices are zero-based.
That is, $\mathcal{B}_{pq}$ is described by the node set as follows:
\begin{equation}
    \mathcal{B}_{pq} \subseteq
    \{0,\ldots,N_\mathrm{x}\}\times\{0,\ldots,N_\mathrm{y}\},
    \label{eq:coarse-bins}
\end{equation}
where $0 \leq p \leq M_\mathrm{x} - 1$ and $0 \leq q \leq M_\mathrm{y} - 1$.
Assuming that each $\mathcal{B}_{pq}$ includes the same number of nodes, let the uniform state $\ket{1_{pq}}$ for each block be defined as
\begin{equation}
    \ket{1_{pq}}
    := \frac{1}{\sqrt{S}}
    \sum_{(v_\mathrm{x}, v_\mathrm{y})\in\mathcal{B}_{pq}} \ket{v_\mathrm{y}, v_\mathrm{x}},
    \label{eq:block-uniform-state}
\end{equation}
where $S := | \mathcal{B}_{pq} |$.
The averaged displacement component on each block is then described as
\begin{align}
    &\overline{u}_{pq, v_\mathrm{c}} \nonumber \\
    &:= \frac{1}{S}
    \sum_{(v_\mathrm{x}, v_\mathrm{y}) \in \mathcal{B}_{pq}} u_{v_\mathrm{x}, v_\mathrm{y}, v_\mathrm{c}} \nonumber \\
    &= \frac{\| \mathbf{u} \|}{\sqrt{S} } (\bra{1_{pq}} \bra{v_c}) \ket{u} \nonumber \\
    &= \frac{ P_\mathrm{max} \| \mathbf{f} \|}{2 \lambda (1 + \nu) \sqrt{S} } \bra{0^{a_+}} \bra{1_{pq}} \bra{v_c} \mathbf{U}_{K_\mathrm{F}^{+}} \ket{0^{a_+}} \ket{f},
    \label{eq:coarse-physical-displacement}
\end{align}
where $u_{v_\mathrm{x}, v_\mathrm{y}, v_\mathrm{c}}$ represents the $v_c$-component of the displacement at the node $(v_\mathrm{x}, v_\mathrm{y})$.
Assuming that the number of nodes in each $B_{pq}$ is power of two, we can prepare the reference state $\ket{1_{pq}}$ by fixing the coarse coordinate bits and applying Hadamard gates to the bits corresponding to the fine coordinate.
In this case, we need to estimate the expected value $\bra{0^{a_+}} \bra{1_{pq}} \bra{v_c} \mathbf{U}_{K_\mathrm{F}^{+}} \ket{0^{a_+}} \ket{f}$.
The number of values to be estimated is
\begin{equation}
    M_{\mathrm{coarse}} = 2 M_\mathrm{x} M_\mathrm{y},
    \label{eq:number-coarse-coefficients}
\end{equation}
which is independent of the number of nodes when the visualization resolution is fixed. 
To estimate the reconstructed displacement field $\bar{\mathbf{u}} := [\bar{u}_{00, 0}, \bar{u}_{00, 1}, \bar{u}_{10, 0}, \ldots, \bar{u}_{M_\mathrm{x}-1 ~ M_\mathrm{y}-1, 1}]^\top$ with accuracy $\epsilon_\mathrm{exp}$, we need to estimate each averaged displacement component $\bar{u}_{pq, v_c}$ with accuracy $\epsilon_\mathrm{exp} / \sqrt{M_\mathrm{coarse}}$, thereby requiring to estimate the expected value within error
\begin{align}
    \frac{2 \lambda (1 + \nu) \sqrt{S} \epsilon_\mathrm{exp}}{ P_\mathrm{max} \| \mathbf{f} \| \sqrt{M_\mathrm{coarse}} }.
\end{align}
Estimating every averaged displacement to this error by amplitude estimation requires
\begin{align}
    \mathcal{O} \left( \frac{ P_\mathrm{max} \| \mathbf{f} \| M_\mathrm{coarse}^{1.5} }{\lambda (1 + \nu) \sqrt{S} \epsilon_\mathrm{exp}} \right) = \mathcal{O} \left( \frac{ P_\mathrm{max} \| \mathbf{f} \| M_\mathrm{coarse}^2 }{\lambda (1 + \nu) \sqrt{N} \epsilon_\mathrm{exp}} \right)
\end{align}
calls to $\mathbf{U}_{K_\mathrm{F}^{+}}$ where we use $M_\mathrm{coarse} S = N$ by definition.

\subsubsection{Space and gate complexity}
\label{subsubsec:global-resource-count}

\begin{table*}[t]
\centering
\caption{Qubit counts, labeled registers, and subnormalization factors for each block-encoding. 
These counts assume the Dirichlet boundary condition imposed on the left edges and the homogeneous material.}
\label{tab:qubit-counts}
\begin{tabular}{llll}
\toprule
Block-encoding & Total qubits & Labeled registers & Normalization \\
\midrule
$U_{\sqrt{e}}$ for $\sqrt{\mathbf{K}_e}$ & $4$ & $a_e, \xi, \eta, c$ & $\sqrt{\lambda (1 + \nu)}$ \\
$U_R$ for $\mathbf{R}$ & $n_\mathrm{x} + n_\mathrm{y} + 6$ & $w, b, \xi, \eta, x, y, c$ & $2$ \\
$O_\mathrm{D}$ for $\mathbf{\Pi}_\mathrm{F}$ & $n_\mathrm{x} + 1$ & $x, a_\mathrm{D}$ & $1$ \\
$U_B$ for $\mathbf{I} - \frac{1}{2\lambda (1 + \nu)}\mathbf{K}_\mathrm{F}$ & $n_\mathrm{x} + n_\mathrm{y} + 8$ & $w, b, \xi, \eta, a_e, a_\mathrm{D}, x, y, c$ & $1$ \\
$U_{K_\mathrm{F}^+}$ for $\mathbf{K}_\mathrm{F}^{+}$ & $n_\mathrm{x} + n_\mathrm{y} + 10$ & $w, b, \xi, \eta, a_e, a_\mathrm{D}, a_\mathrm{L}, a_s, x, y, c$ & $\frac{P_\mathrm{max}}{2\lambda (1 + \nu)}$ \\
\bottomrule
\end{tabular}
\end{table*}

\begin{table*}[t]
\centering
\caption{Gate counts for each quantum circuit module.}
\label{tab:gate-counts}
\begin{tabular}{lrrrrrr}
\hline
Module & $\mathbf{H}$ & $\mathbf{X}$ & CNOT & $\mathbf{T}$ & $\mathbf{S}$ & $\mathbf{R}_y$ \\
\hline

$\mathbf{U}_{\sqrt{e}}$
& $26$ & -- & $46$ & $32$ & $4$ & $16$ \\

Controlled-$\mathbf{U}_{\sqrt{e}}$
& $72$ & -- & $122$ & $118$ & $16$ & $32$ \\

$\mathbf{U}_{\mathrm{flag}}$
& $16n-14$ & $10n-28$ & $24n+8$ & $32n-1$ & -- & -- \\

$\mathbf{U}_R$
& $22n$ & $10n-28+\mathcal{O}(n)$ & $24n+8+\mathcal{O}(n)$
& $53n+41$ & -- & -- \\

$\mathbf{U}_L$
& $22n+72$ & $10n-26+\mathcal{O}(n)$ & $24n+130+\mathcal{O}(n)$
& $53n+159$ & $16$ & $32$ \\

$\mathbf{R}_{L,\mathrm{out}}$
& $12$ & $12$ & $18$ & $23$ & -- & -- \\

$\mathbf{U}_B$
& $44n+156$ & $20n-40+\mathcal{O}(n)$ & $48n+278+\mathcal{O}(n)$
& $106n+341$ & $32$ & $64$ \\

\hline
\end{tabular}
\par\smallskip
\footnotesize
Here, $n=n_x+n_y$, where $n_x=\lceil\log_2(N_x+1)\rceil$ and $n_y=\lceil\log_2(N_y+1)\rceil$.
The costs of implementing oracles for boundary conditions and arbitrary rotation synthesis are excluded.
\end{table*}

Here, we derive the space and gate complexity for solving the linear system $\mathbf{K} \mathbf{u} = \mathbf{f}$.
The principal qubit counts and normalization constants for the explicit construction of each block-encoding are summarized in Table~\ref{tab:qubit-counts}. 
The system register corresponding to the node indices contains $n_\mathrm{x} + n_\mathrm{y} + 1$ qubits. 
The normalized block-encoding of the shifted stiffness matrix uses seven ancilla qubits: one membership flag qubit $a_\mathrm{D}$, two working qubits $w$, one invalid flag qubit $b$, two local label qubits $(\xi, \eta)$, and one local element block-encoding ancilla $a_e$.
The block-encoding of the pseudoinverse of the stiffness matrix uses nine ancilla qubits where one ancilla qubit $a_\mathrm{L}$ for LCU and another qubit $a_s$ for QSVT are required in addition to those for $\mathbf{U}_B$.
Thus, the qubit count is $\mathcal{O}(\log N)$ with a constant ancilla overhead.
The table does not include temporary workspace used inside multi-controlled gates, decrementers, material oracles, or load state preparation.
These ancilla qubits are at most iterated-logarithmic in the sizes of coordinate registers and can be reused sequentially.

Next, we estimate the gate complexity of our framework.
The gate counts for implementing each block-encoding and primitive are derived in Appendix~\ref{sec:gate_count} and are summarized in Table~\ref{tab:gate-counts}.
When $\mathbf{R}_\mathrm{y}$ rotations and QSVT phase rotations are synthesized over the Clifford+T gate set, their errors must be taken into account, which yields additional logarithmic factors $\log (1/\epsilon)$ in the $\mathbf{T}$ count.
As discussed in Section~\ref{subsubsec:normalized-B-reflection} and listed in Table~\ref{tab:gate-counts}, the normalized block-encoding $\mathbf{U}_B$ includes $\mathcal{O}(\log N)$ Clifford+T gates.
Since the number of calls to $\mathbf{U}_B$ or $\mathbf{U}_B^\dagger$ in the linear system solver is $\mathcal{O}(\sqrt{\kappa} \log (\kappa / (\lambda (1+\nu) \epsilon)))$, the gate complexity for preparing the displacement state is $\mathcal{O}(\sqrt{\kappa} \log N \log (\kappa / (\lambda (1+\nu) \epsilon)))$.
Considering the additional overhead for the amplitude amplification, we estimate the gate complexity of our framework as 
\begin{align}
    \mathcal{O} \left( \frac{\| \mathbf{f} \| }{ \| \mathbf{u} \| } \kappa^{1.5} \log N \log \frac{\kappa} { \lambda (1+\nu) \epsilon} \right),
\end{align}
where the scaling of $\| \mathbf{f} \|$ depends on the force types and $\kappa \in \Theta(N)$.
When considering the specific observables as discussed in the previous section, additional overhead for amplitude estimation is required.
The gate complexity for specific tasks are summarized in Table~\ref{tab:complexity}, where we include the complexities of classical counterparts: a direct method by nested dissection~\cite{george1973nested} and a conjugate gradient method~\cite{saad2003iterative}. 
As shown in Table~\ref{tab:complexity}, the case with body forces is the most favorable owing to the constant success probability of postselection.
The resulting costs scale as $\widetilde{\mathcal{O}}(N^{1/2})$ under the assumption of $\kappa \in \Theta(N)$ in two dimensions. 
This leaves room for a polynomial quantum speedup relative to sparse direct methods and conjugate gradient method. 
Note that the table does not include the cost of state preparation for input loading. 
This cost can become a bottleneck and potentially offset the quantum speedup for general input states.
However, efficient or approximate state preparation methods have been developed for states with suitable structures, including approaches based on tensor networks and parameterized quantum circuits~\cite{nakaji2022approximate, melnikov2023quantum, shirakawa2024automatic}.

\begin{table*}[t]
\centering
\caption{Space and gate complexities for each task: displacement state preparation (DSP), strain energy estimation (SEE), and low-resolution tomography (LRT).
The gate complexity depends on the type of applied forces.
The notation $\tilde{\mathcal{O}}$ suppresses the logarithmic factors. CG represents the conjugate gradient.}
\label{tab:complexity}
\begin{tabular}{m{.9\columnwidth}m{0.35\columnwidth}m{0.6\columnwidth}}
\toprule
Tasks & Space complexity & Gate complexity \\
\midrule
DSP, SEE, LRT by direct method (classical) & $\mathcal{O}(N)$ & $\mathcal{O}(N^{1.5})$ \\
DSP, SEE, LRT by CG method (classical) & $\mathcal{O}(N)$ & $\mathcal{O}(N \sqrt{\kappa} \log \frac{1}{\epsilon}) $ \\
DSP with body force (This work) & $\log N$ & $\mathcal{O} \left( \frac{\kappa^{1.5}}{N} \log N \log \frac{\kappa} { \lambda (1+\nu) \epsilon} \right)$  \\
DSP with boundary traction (This work) & $\log N$ & $\mathcal{O} \left( \frac{\kappa^{1.5}}{N^{0.75}} \log N \log \frac{\kappa} { \lambda (1+\nu) \epsilon} \right)$  \\
DSP with point force (This work) & $\log N$ & $\mathcal{O} \left( \frac{\kappa^{1.5}}{N^{0.5}} \log N \log \frac{\kappa} { \lambda (1+\nu) \epsilon} \right)$  \\
SEE with body force (This work) & $\log N$ & $\tilde{\mathcal{O}} \left( \frac{ \kappa^{1.5} \log N }{N \lambda(1+\nu) \epsilon_\mathrm{exp}} \right)$  \\
SEE with boundary traction (This work) & $\log N$ & $\tilde{\mathcal{O}} \left( \frac{ \kappa^{1.5} \log N }{ N^{0.5} \lambda(1+\nu) \epsilon_\mathrm{exp}} \right)$  \\
SEE with point force (This work) & $\log N$ & $\tilde{\mathcal{O}} \left( \frac{ \kappa^{1.5} \log N }{\lambda(1+\nu) \epsilon_\mathrm{exp}} \right)$  \\
LRT with body force (This work) & $\log N$ & $\tilde{\mathcal{O}} \left( \frac{ M_\mathrm{coarse}^2 \kappa^{1.5} \log N }{ N \lambda (1 + \nu)  \epsilon_\mathrm{exp}} \right)$  \\
LRT with boundary traction (This work) & $\log N$ & $\tilde{\mathcal{O}} \left( \frac{ M_\mathrm{coarse}^2 \kappa^{1.5} \log N }{ N^{0.75} \lambda (1 + \nu)  \epsilon_\mathrm{exp}} \right)$  \\
LRT with point force (This work) & $\log N$ & $\tilde{\mathcal{O}} \left( \frac{ M_\mathrm{coarse}^2 \kappa^{1.5} \log N  }{ N^{0.5} \lambda (1 + \nu)  \epsilon_\mathrm{exp}} \right)$  \\
\bottomrule
\end{tabular}
\end{table*}

Although the explicit circuit construction and complexity analysis presented above focus on two-dimensional problems, the proposed framework can be extended to three dimensions.
For trilinear hexahedral elements, each element has 8 nodes and 24 displacement degrees of freedom, independent of the global mesh size.
Each global node belongs to at most 8 elements, and the decrementers for three coordinates require only $O(\log N)$ gates.
Therefore, the normalized block-encoding of the shifted stiffness matrix maintains an $O(\log N)$ gate complexity.
As discussed in Section~\ref{subsec:element-block-encoding}, the element stiffness matrix scales as $\mathbf{K}_e^{(3)} = Eh\widehat{\mathbf{K}}_e^{(3)}$.
Accordingly, the normalization for the stiffness matrix satisfies $\alpha_K \in \Theta(h)$, yielding
$\beta = 2/\alpha_K\in\Theta(h^{-1})$.
Since $\varsigma_{\min}^{+}(\mathbf{K}_F)\in\Theta(h^3)$,
we obtain $\kappa_B^\star\in\Theta(h^{-2})=\Theta(N^{2/3})$.
Thus, the normalization for the pseudoinverse scales as
$\alpha_{K_F^+}=\beta P_{\max} \in \Theta(N)$,
Consequently, a calculation similar to that for the two-dimensional case yields the gate complexities for displacement state preparation as
$\widetilde{O}(N^{1/3})$, $\widetilde{O}(N^{1/2})$,
and $\widetilde{O}(N^{5/6})$, respectively.
Hence, the dependency on the degrees of freedom $N$  becomes more favorable in three dimensions, which is consistent with the general observation that quantum FEM becomes more favorable as the spatial dimension increases~\cite{montanaro2016quantum}.

\section{Numerical Experiments} \label{sec:results}

To validate the proposed method, we performed numerical simulations of the deformation of a cantilever beam in two dimensions. 
Specifically, we considered a rectangular domain with its left boundary fixed and a downward point load applied at its lower-right corner. 
The quantum circuits were implemented using Qiskit~\cite{javadi2024quantum}.
The proposed quantum linear system solver requires the parameter $\kappa_B$, which characterizes the inverse of the smallest nonzero eigenvalue of the scaled stiffness matrix and is of the same asymptotic order as its condition number. 
In general, this parameter is not known a priori. 
In the present numerical experiments, we use values of $\kappa_B$ determined by classical computation to focus on validating the correctness of the proposed quantum framework. 
In practical applications, $\kappa_B$ can instead be estimated, for example, by extrapolating the condition numbers of smaller finite element systems, as discussed in Appendix~\ref{app:kappa-estimation}.

First, to verify the implementation of the normalized block encoding $U_B$, we evaluated the following max norm using $U_B$ evaluated classically with a unitary simulator and the classically assembled stiffness matrix $\mathbf{K}_{F,\mathrm{classical}}$:
\begin{align}
\left| \Pi_{L,\mathrm{in}}U_B\Pi_{L,\mathrm{in}}
- \left( I-\frac{\mathbf{K}_{F,\mathrm{classical}}}{2\lambda(1+\nu)} \right) \right|_\mathrm{max}.
\end{align}
For $N_x=N_y=3$ and $\nu=0.3$, the max norm was less than $10^{-14}$, confirming that the proposed quantum circuit correctly block-encodes the stiffness matrix.

\begin{figure}[t]
\centering
\subfloat[][]{\includegraphics[width=\columnwidth]{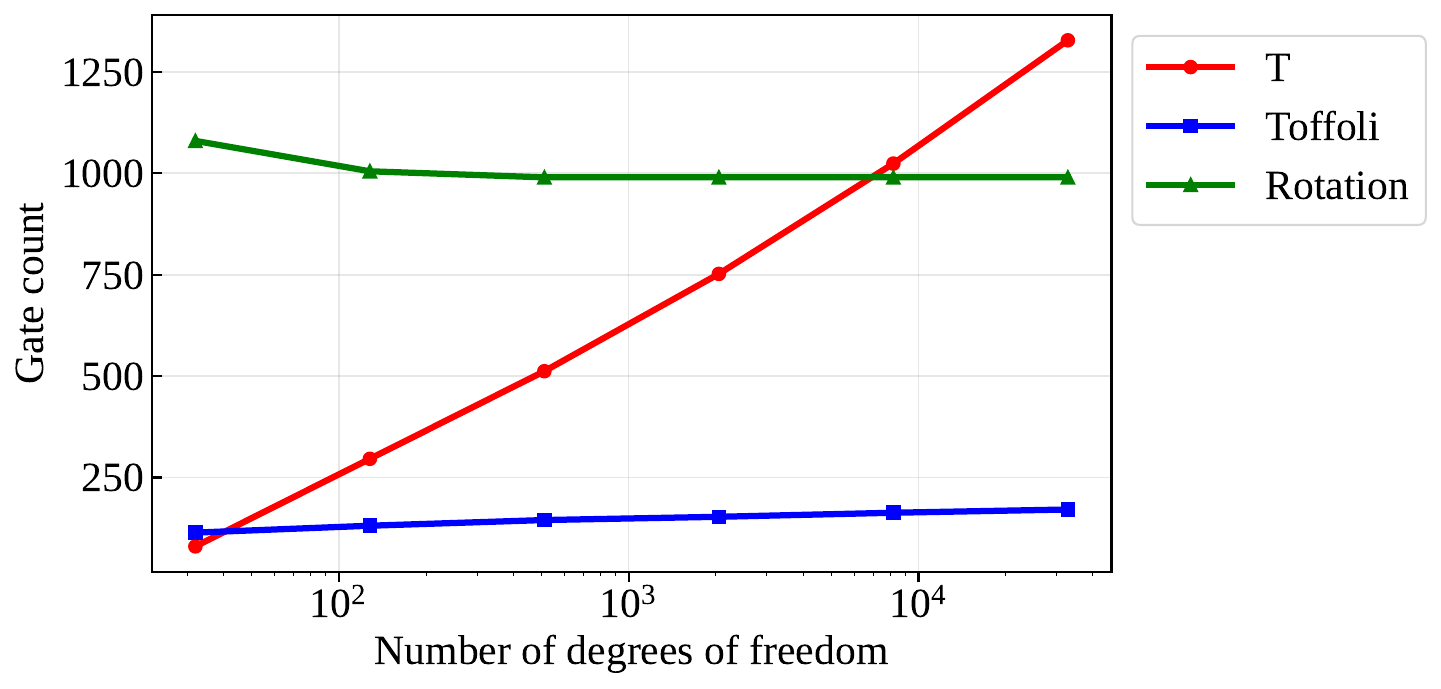}}\\
\subfloat[][]{\includegraphics[width=\columnwidth]{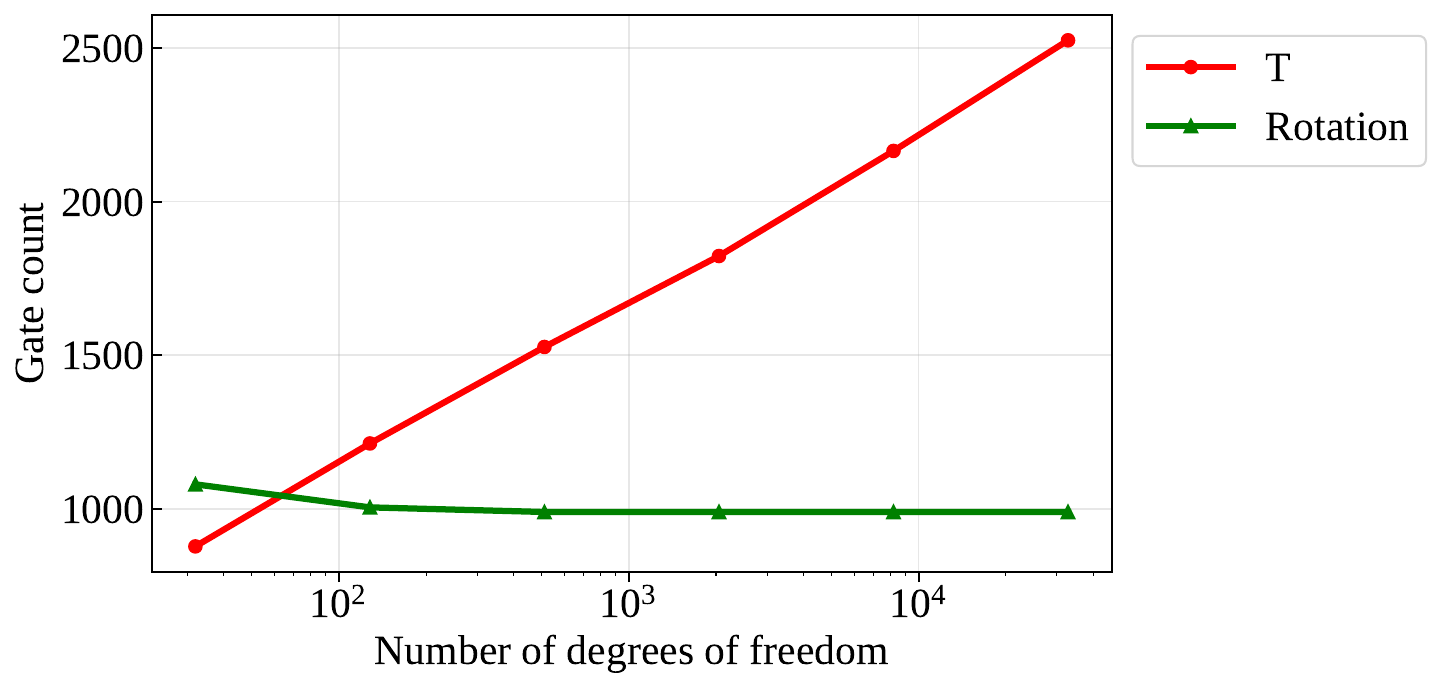}}
\caption{Gate counts of the quantum circuit implementing the normalized block encoding $U_B$. (a) Toffoli gates, $\mathbf{T}$ gates, and arbitrary single-qubit rotation gates are included in the gate basis as non-Clifford gates. (b) $\mathbf{T}$ gates and arbitrary single-qubit rotation gates are included in the gate basis as non-Clifford gates, with Toffoli gates decomposed into Clifford+T gates. We set $N_x=N_y$.}
\label{fig:gatecounts}
\end{figure}

\begin{figure}[t]
\centering
\subfloat[][]{\includegraphics[width=\columnwidth]{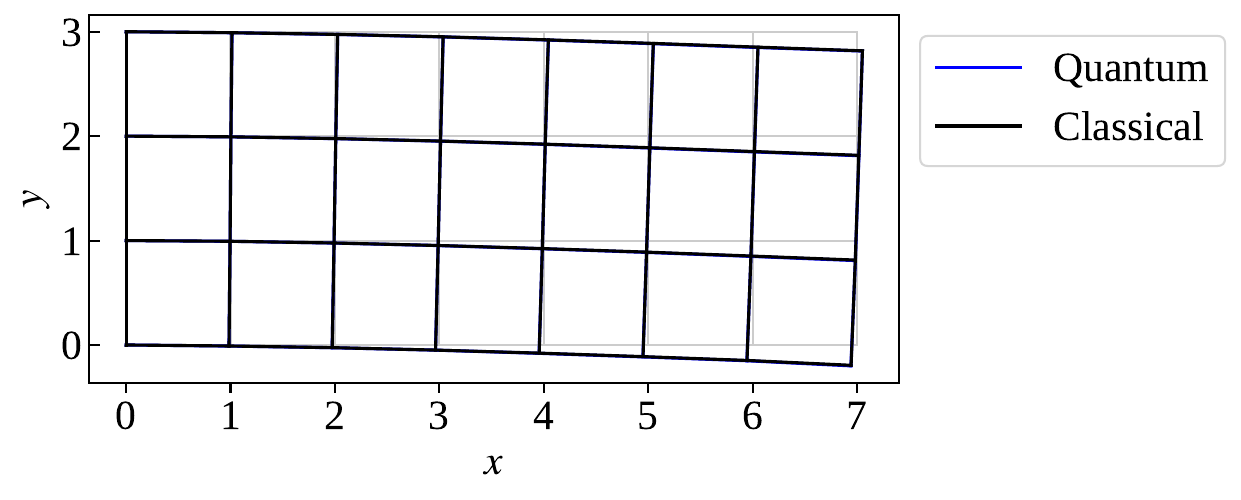}} \\
\subfloat[][]{\includegraphics[width=\columnwidth]{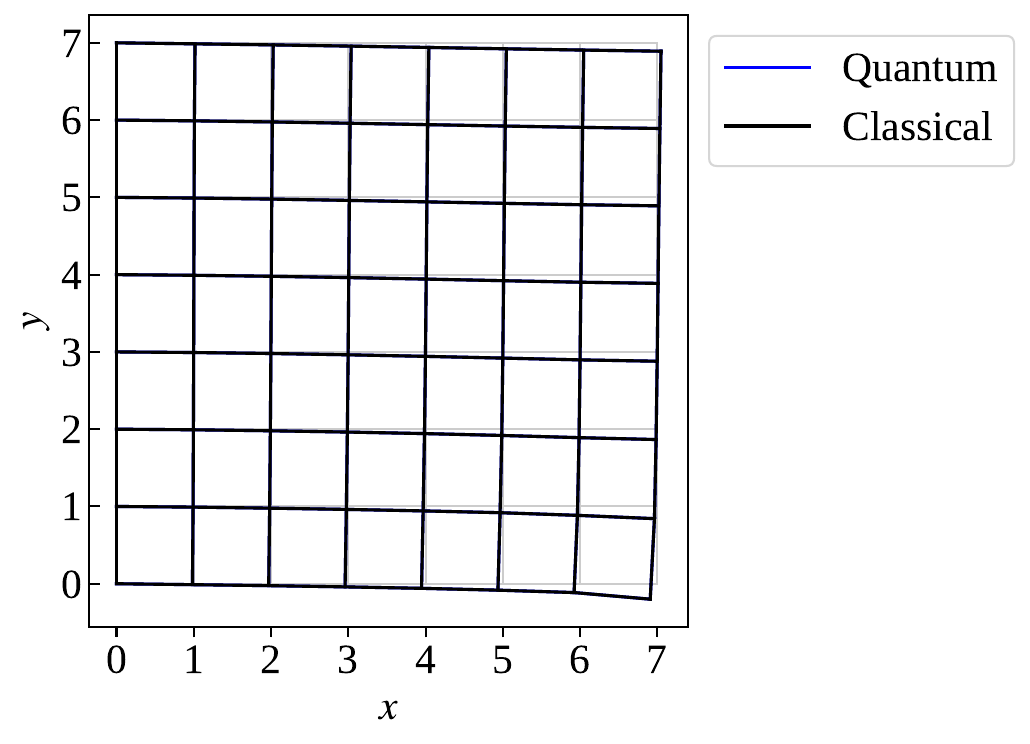}}
\caption{Results of deformation simulations of the cantilever beam for (a) a $7\times3$ mesh and (b) a $7\times7$ mesh.}
\label{fig:result}
\end{figure}

Next, we investigated the gate count of the quantum circuit implementing the normalized block encoding $U_B$. Figure~\ref{fig:gatecounts} shows the gate counts for two gate bases: one in which Toffoli gates, $\mathbf{T}$ gates, and arbitrary single-qubit rotation gates are included as non-Clifford gates, and another in which the Toffoli gates are decomposed into Clifford+T gates. The gate counts were obtained by decomposing the quantum circuit for $U_B$ using the Qiskit transpiler. The horizontal axis represents the number of degrees of freedom $N$ of the stiffness matrix on a logarithmic scale. The number of degrees of freedom was varied while maintaining $N_x=N_y$.

As shown in Fig.~\ref{fig:gatecounts}, the number of arbitrary single-qubit rotation gates remains nearly constant with respect to the number of degrees of freedom. This is because these gates arise primarily from the controlled rotations in Eq.~\eqref{eq:multiplexed-ry-angle}, which are used in the block encoding of the square root of the element stiffness matrix. The number of gates required for these controlled rotations is independent of the mesh size.
The number of arbitrary single-qubit rotation gates is also observed to be slightly larger for systems with fewer degrees of freedom. In addition to the controlled rotations described above, single-qubit rotation gates arise from the decomposition of multi-controlled NOT gates. The decomposition of a multi-controlled NOT gate depends on the number of available ancilla qubits, including other qubits in the circuit that can be used as dirty ancillas. Consequently, as the number of qubits increases, the transpiler can exploit more efficient circuit decompositions, which likely accounts for the observed reduction in the number of arbitrary single-qubit rotation gates.

Even when arbitrary single-qubit rotation gates are decomposed into $\mathbf{T}$ gates, each such rotation requires only a constant number of $\mathbf{T}$ gates for a fixed approximation accuracy. 
Moreover, the numbers of Toffoli and $\mathbf{T}$ gates increase only logarithmically with the number of degrees of freedom of the stiffness matrix. 
These results demonstrate that the proposed method enables an efficient implementation of the block encoding.

Finally, using a statevector simulator, we performed quantum circuit simulations of the cantilever-beam deformation problem, including the solution of the linear system. The results are shown in Fig.~\ref{fig:result}, together with those obtained by classical computation. For the classical computation, the linear system was solved using \texttt{numpy.linalg.solve}. As shown in Fig.~\ref{fig:result}, for both the $7\times3$ and $7\times7$ meshes, the solutions obtained using the proposed method are in good agreement with the corresponding classical solutions.

\section{Conclusions} \label{sec:conclusions}
In this work, we considered isotropic linear elastic materials discretized on a uniform two-dimensional mesh and developed a block encoding of the stiffness matrix that can be implemented with a number of gates that scales logarithmically with the number of finite elements. 
We further presented a framework for preparing a quantum state whose amplitudes encode the displacement vector by applying a quantum linear system algorithm to the resulting block encoding.

Although the computational cost of the proposed method was shown to exhibit more favorable scaling than that of representative classical methods, whether a quantum speedup can be maintained for the overall framework when the overhead associated with quantum measurements is taken into account requires further careful investigation. 
Since the computational cost of the proposed method is dominated by its dependence on the condition number of the stiffness matrix, incorporating preconditioning techniques to reduce the condition number is also an important direction for future work. 
Classical multilevel preconditioners of Bramble--Pasciak--Xu type are designed to make the condition number of elliptic finite element systems independent of mesh refinement~\cite{bramble1990parallel}. 
Deiml and Peterseim have shown that a BPX hierarchy can be implemented on a quantum computer for scalar elliptic PDEs and can restore favorable end-to-end complexity~\cite{deiml2024quantum}. 
Extending this strategy to isotropic elasticity is an important next step. 
If a mesh-independent effective condition number is achieved, the inverse-polynomial degree becomes polylogarithmic, leaving state preparation, observable estimation, and discretization accuracy as the principal remaining costs.

\begin{appendices}

\section*{Appendix}

\makeatletter
\def\@seccntformat#1{%
  \csname the#1\endcsname\quad}
\makeatother

\section{Gate counts} \label{sec:gate_count}

Here, we provide the gate counts required to implement each block-encoding constructed in the main text.
We employ the gate set $\{\mathbf{H}, \mathbf{S}, \mathbf{X}, \mathrm{CNOT}, \mathbf{T}, \mathbf{R}_y \}$, consisting of single-qubit gates and the CNOT gate.
Note that the gate set $\{\mathbf{H}, \mathbf{S}, \mathrm{CNOT}, \mathbf{T}\}$ is universal.
The $\mathbf{X}$ gate can be decomposed as $\mathbf{X} = \mathbf{H}\mathbf{S}^2\mathbf{H}$, and arbitrary $\mathbf{R}_y$ rotations can be approximated to any desired accuracy using Clifford+T gates.
Nevertheless, we count $\mathbf{X}$ gates separately, as they can be implemented natively on many quantum computers.
We also count $\mathbf{R}_y$ gates separately to characterize the gate complexity of the circuits before Clifford+T synthesis, thereby avoiding additional approximation errors associated with the synthesis of arbitrary rotations.

\subsection{Block-encoding of the square root of the element stiffness matrix} \label{sec:count_sqrt_Ke}

The implementation of $\mathbf{U}_{\sqrt{e}}$ uses six Hadamard gates, two CNOT gates, two controlled-$\mathbf{U}_{\eta, c}$ gates, and one uniformly controlled $\mathbf{R}_\mathrm{y}$ gate $\mathbf{U}_{\sqrt{D}}$. 
The controlled-$\mathbf{U}_{\eta, c}$ gates includes one controlled Hadamard gate, one controlled $\mathbf{Z}$ gate, two controlled $\mathbf{R}_\mathrm{y}$ gates and two Toffoli gates as suggested by Eq.~\eqref{eq:U4}.
The controlled Hadamard gates, denoted by $C(\mathbf{H})$, can be decomposed into
\begin{align}
    C(\mathbf{H}) = (\mathbf{I} \otimes (\mathbf{H} \mathbf{S} \mathbf{H} \mathbf{T}) ) C(\mathbf{X}) (\mathbf{I} \otimes (\mathbf{T}^\dagger \mathbf{H} \mathbf{S}^\dagger \mathbf{H} ) ),
\end{align}
which includes one CNOT gate, two $\mathbf{T}$ gates, four Hadamard gates, and two $\mathbf{S}$ gates.
Throughout this paper, we count $\mathbf{T}^\dagger$ gates as $\mathbf{T}$ gates.
The controlled $\mathbf{Z}$ gate can be decomposed into one CNOT and two Hadamard gates.
One controlled $\mathbf{R}_\mathrm{y}$ gate is realized by two CNOT and two $\mathbf{R}_\mathrm{y}$ gates, and one Toffoli gate can be decomposed into six CNOT gates, seven $\mathbf{T}/\mathbf{T}^\dagger$ gates, and two Hadamard gates.
Thus, the controlled-$\mathbf{U}_{\eta, c}$ is implemented using 10 Hadamard gates, two $\mathbf{S}$ gates, 18 CNOT gates, 16 $\mathbf{T}$ gates, and 4 $\mathbf{R}_\mathrm{y}$ gates.
A uniformly three-qubit-controlled $\mathbf{R}_\mathrm{y}$ gate can be implemented using $8(=2^3)$ CNOT and $8(=2^3)$ $\mathbf{R}_\mathrm{y}$ gates~\cite{mottonen2004quantum}.
Consequently, one application of $U_{\sqrt{e}}$ requires 
\begin{itemize}
    \item 26 Hadamard gates;
    \item 4 $\mathbf{S}$ gates;
    \item 46 CNOT gates;
    \item 32 $\mathbf{T}$ gates;
    \item and 16 $\mathbf{R}_\mathrm{y}$ gates.
\end{itemize}

In the finite element assembly, we use the controlled version of $\mathbf{U}_{\sqrt{e}}$, which requires six controlled Hadamard gates, two Toffoli gates, two controlled-controlled-$\mathbf{U}_{\eta, c}$ gates, and one controlled version of the uniformly controlled $\mathbf{R}_\mathrm{y}$ gate $\mathbf{U}_{\sqrt{D}}$.
The controlled Hadamard gates includes one CNOT gate, two $\mathbf{T}$ gates, four Hadamard gates, and two $\mathbf{S}$ gates.
The controlled-controlled-$\mathbf{U}_{\eta, c}$ gate includes one controlled-controlled Hadamard gate, one controlled-controlled $\mathbf{Z}$ gate, two controlled-controlled $\mathbf{R}_\mathrm{y}$ gates, and two 3-controlled $\mathbf{X}$ gates.
One controlled-controlled Hadamard gate is decomposed into one Toffoli gate, two $\mathbf{T}$ gates, four Hadamard gate, and two $\mathbf{S}$ gates, and with the Toffoli gate decomposed, it yields nine $\mathbf{T}$ gates, six Hadamard gates, two $\mathbf{S}$ gates, and six CNOT gates.
One controlled-controlled $\mathbf{Z}$ gate includes one Toffoli gate and two Hadamard gates, which results in seven $\mathbf{T}$, four Hadamard, and six CNOT gates.
One controlled-controlled $\mathbf{R}_\mathrm{y}$ gate is decomposed into four CNOT and four $\mathbf{R}_\mathrm{y}$ gates.
One 3-controlled $\mathbf{X}$ gate is decomposed into two relative phase Toffoli gates and one Toffoli gate with one clean ancilla qubit, which results in 12 CNOT, 15 $\mathbf{T}$, and six Hadamard gates.
Hence, the controlled-controlled-$\mathbf{U}_{\eta, c}$ gate is implemented using 22 Hadamard, two $\mathbf{S}$, 46 $\mathbf{T}$, 8 $\mathbf{R}_\mathrm{y}$, and 44 CNOT gates.
One controlled version of the uniformly controlled $\mathbf{R}_\mathrm{y}$ gate $\mathbf{U}_{\sqrt{D}}$ can be implemented using $16(=2^4)$ CNOT and $16(=2^4)$ $\mathbf{R}_\mathrm{y}$ gates~\cite{mottonen2004quantum}.
Consequently, one application of the controlled version of $\mathbf{U}_{\sqrt{e}}$ requires 
\begin{itemize}
    \item 72 Hadamard gates; %
    \item 16 $\mathbf{S}$ gates; %
    \item 122 CNOT gates; %
    \item 118 $\mathbf{T}$ gates; %
    \item and 32 $\mathbf{R}_\mathrm{y}$ gates; %
\end{itemize}

\subsection{Block-encoding of the element restriction operator} \label{sec:count_restriction}

The implementation of $\mathbf{U}_R$ uses two Hadamard gates, one $\mathbf{U}_\mathrm{flag}$, and one $\mathbf{U}_\mathrm{decr}$.
One application of $\mathbf{U}_\mathrm{flag}$ requires four $(n_\mathrm{x} + 1)$-controlled Toffoli gates, four $(n_\mathrm{y} + 1)$-controlled Toffoli gates, one Toffoli gate, two CNOT gates, and $2(n_\mathrm{x} + 1) + 2(n_\mathrm{y} + 1)$ $\mathbf{X}$ gates, where $\mathbf{X}$ gates are required for flipping the control bits to the $\ket{1}$ state.
A $k(\geq 3)$-controlled Toffoli gate can be decomposed into $(2k-4)$ relative phase Toffoli gates, one Toffoli gate, and $2k-6$ $\mathbf{X}$ gates, with one clean ancilla qubit~\cite{khattar2025rise}.
A relative phase Toffoli gate can be implemented using three CNOT and four $\mathbf{T}$ gates, and two Hadamard gates, while the Toffoli gate requires six CNOT, seven $\mathbf{T}$ gates, and two Hadamard gates~\cite{maslov2016advantage}.
Thus, $k(\geq 3)$-controlled Toffoli gate can be implemented using $6k-6$ CNOT, 
$8k-9$ $\mathbf{T}$ gates, $4k-6$ Hadamard gates, and $2k-6$ $\mathbf{X}$ gates.
Consequently, one application of $\mathbf{U}_\mathrm{flag}$ requires 
\begin{itemize}
    \item $16(n_\mathrm{x} + n_\mathrm{y}) - 14$ Hadamard gates;
    \item $10(n_\mathrm{x} + n_\mathrm{y})-28$ $\mathbf{X}$ gates;
    \item $24 (n_\mathrm{x} + n_\mathrm{y}) + 8$ CNOT gates;
    \item and $32(n_\mathrm{x} + n_\mathrm{y}) -1$ $\mathbf{T}$ gates.
\end{itemize}

A controlled $k$-qubit decrementer can be constructed by regarding the control qubit as an additional least significant bit of an $(k+1)$-qubit decrementer. 
The naive implementation, based on the standard multi-controlled Toffoli ladder circuit, consists of one CNOT gate, one Toffoli gate, and one $j$-controlled Toffoli gate for each $j=3, \ldots, k$. 
Using the decomposition described above for each multi-controlled Toffoli gate~\cite{khattar2025rise,maslov2016advantage}, the required gate counts are
\begin{itemize}
    \item $2+\sum_{j=3}^{k}(4j-6)=2k^2-4k+2$ Hadamard gates;
    \item $\sum_{j=3}^{n}(2j-6)=(k-2)(k-3)$ $\mathbf{X}$ gates;
    \item $1+6+\sum_{j=3}^{k}(6j-6)=3k^2-3k+1$ CNOT gates;
    \item and $7+\sum_{j=3}^{k}(8j-9)=4k^2-5k+1$ $\mathbf{T}$ gates.
\end{itemize}
A single clean ancilla qubit can be reused throughout the multi-controlled Toffoli ladder. 
These counts assume that each multi-controlled Toffoli gate is decomposed independently and that no additional gate cancellations between adjacent decompositions are exploited.
Since $\mathbf{U}_\mathrm{decr}$ consists of $\xi$-controlled $n_\mathrm{x}$-qubit decrementer and $\eta$-controlled $n_\mathrm{y}$-qubit decrementer, the naive implementation of the decrementer makes $\mathbf{U}_\mathrm{decr}$ require $\mathcal{O}(n_\mathrm{x}^2 + n_\mathrm{y}^2)$ CNOT, $\mathbf{T}$, Hadamard, and $\mathbf{X}$ gates.
This dependency can be improved by using more dedicated construction of the decrementer.

Khattar and Gidney showed that a $k$-qubit incrementer can be implemented using $3k$ Toffoli gates and $\log_{2}^{*}k$ clean ancilla qubits~\cite{khattar2025rise} where $\log_{2}^{*}k$ denotes the iterated logarithm. 
The construction additionally contains $\mathcal{O}(k)$ CNOT and $\mathbf{X}$ gates. 
Since a decrementer is the inverse of an incrementer, a controlled $k$-qubit decrementer can be implemented by $3(k+1)$ Toffoli gates, $\mathcal{O}(k)$ CNOT gates, $\mathcal{O}(k)$ $\mathbf{X}$ gates, and $\log_{2}^{*}(k+1)$ clean ancilla qubits. 
Consequently, one application of $\mathbf{U}_R$ requires
\begin{itemize}
    \item $22(n_\mathrm{x} + n_\mathrm{y})$ Hadamard gates;
    \item $10(n_\mathrm{x} + n_\mathrm{y}) - 28 + \mathcal{O}(n_\mathrm{x} + n_\mathrm{y})$ $\mathbf{X}$ gates;
    \item $24(n_\mathrm{x} + n_\mathrm{y}) + 8 + \mathcal{O}(n_\mathrm{x} + n_\mathrm{y})$ CNOT gates;
    \item and $53(n_\mathrm{x} + n_\mathrm{y}) + 41$ $\mathbf{T}$ gates.
\end{itemize}

\subsection{Block-encoding of a Gram factor} \label{sec:count_gram}

One application of $\mathbf{U}_L$ consists of the membership oracle $\mathbf{O}_{\mathrm{D}}$, the block-encoding of the restriction operator $\mathbf{U}_R$, and the block-encoding of the square root of the element stiffness matrix $\mathbf{U}_{\sqrt{e}}$ controlled by the zero state of $b$, which corresponds to the valid branch.
Consequently, one application of $\mathbf{U}_L$ uses
\begin{itemize}
    \item $22(n_\mathrm{x} + n_\mathrm{y}) + 72$ Hadamard gates;
    \item $16$ $\mathbf{S}$ gates;
    \item $10(n_\mathrm{x} + n_\mathrm{y}) - 26 + \mathcal{O}(n_\mathrm{x} + n_\mathrm{y})$ $\mathbf{X}$ gates;
    \item $24(n_\mathrm{x} + n_\mathrm{y}) + 130 + \mathcal{O}(n_\mathrm{x} + n_\mathrm{y})$ CNOT gates;
    \item $53(n_\mathrm{x} + n_\mathrm{y}) + 159$ $\mathbf{T}$ gates;
    \item and 32 $\mathbf{R}_\mathrm{y}$ gates,
\end{itemize}
where we omit the cost of membership oracle $\mathbf{O}_\mathrm{D}$ because it depends on the boundary conditions.

\subsection{Normalized block-encoding of the shifted stiffness matrix} \label{sec:count_reflection}

The output reflection $\mathbf{R}_{L, \mathrm{out}}$ acts only on registers $(a_\mathrm{D},b,w,a_e)$ and is therefore implemented by a 4-controlled $\mathbf{Z}$ and twice applications of five $\mathbf{X}$ gates flipping the zero state to the one state.
The 4-controlled $\mathbf{Z}$ gate can be transformed into two Hadamard gates and one 4-controlled $\mathbf{X}$ gate, which can be decomposed into four relative phase Toffoli gates, one Toffoli gate and two $\mathbf{X}$ gates, with one clean ancilla qubit~\cite{khattar2025rise}.
Hence, the output reflection is implemented using 12 Hadamard, 12 $\mathbf{X}$, 18 CNOT and 23 $\mathbf{T}$ gates.
Therefore, the normalized block-encoding of the shifted stiffness matrix is implemented using
\begin{itemize}
    \item $44(n_\mathrm{x} + n_\mathrm{y}) + 156$ Hadamard gates;
    \item $32$ $\mathbf{S}$ gates;
    \item $20(n_\mathrm{x} + n_\mathrm{y}) - 40 + \mathcal{O}(n_\mathrm{x} + n_\mathrm{y})$ $\mathbf{X}$ gates;
    \item $48(n_\mathrm{x} + n_\mathrm{y}) + 278 + \mathcal{O}(n_\mathrm{x} + n_\mathrm{y})$ CNOT gates;
    \item $106(n_\mathrm{x} + n_\mathrm{y}) + 341$ $\mathbf{T}$ gates;
    \item and 64 $\mathbf{R}_\mathrm{y}$ gates.
\end{itemize}

\section{Block-encoding of stiffness matrix for spatially varying Young's modulus}
\label{sec:heterogeneous-material}

\setcounter{equation}{0}

Here, we discuss the extension of our construction to materials with spatially varying Young's modulus.
We assume that the Poisson ratio is constant while the Young's modulus is element-wise constant. 
Let us choose a reference modulus $E_{\mathrm{ref}}>0$ and define
\begin{equation}
    \mathbf{D}_{\rho} := \texttt{diag}(\rho_{00},\rho_{10},\ldots,\rho_{N_\mathrm{x}-1,N_\mathrm{y}-1}),
    \label{eq:material-diagonal}
\end{equation}
with
\begin{align}
    \rho_{e_\mathrm{x} e_\mathrm{y}} := \frac{E_{e_\mathrm{x} e_\mathrm{y}}}{E_{\mathrm{ref}}}.
\end{align}
Let $\lambda_{\mathrm{ref}}$ be the reference parameter $E_{\mathrm{ref}}/(1 - \nu^2)$. 
The global stiffness matrix  can then be constructed by
\begin{equation}
    \mathbf{K}_{\rho}=\lambda_{\mathrm{ref}}
    \mathbf{R}^{\top}
    \left(\mathbf{D}_{\rho} \otimes \widehat{\mathbf{K}}_e \right)
    \mathbf{R},
    \label{eq:heterogeneous-assembly}
\end{equation}
which is an extension of Eq.~\eqref{eq:assembled-homogeneous}.
Consequently, we need to replace $\mathbf{I}^{(N_\mathrm{el})}$ by $\sqrt{\mathbf{D}_{\rho}}$ in Eq.~\eqref{eq:gram-factor-L}.

There are several ways to implement the block-encoding of $\sqrt{\mathbf{D}_{\rho}}$.
An example is to assume a bit oracle defined by
\begin{equation}
    \mathbf{O}_{\rho}\ket{e_\mathrm{x}, e_\mathrm{y}} \ket{0^q}
    =\ket{e_\mathrm{x}, e_\mathrm{y}}\ket{\widetilde{\sqrt{\rho}}_{e_\mathrm{x} e_\mathrm{y}}},
    \label{eq:rho-value-oracle}
\end{equation}
where $\widetilde{\sqrt{\rho}}_{e_\mathrm{x} e_\mathrm{y}}$ is a fixed-point approximation of $\sqrt{\rho}_{e_\mathrm{x} e_\mathrm{y}}$. 
A controlled rotation
\begin{equation}
    \mathbf{R}_\mathrm{y} \! \left(2\arccos \sqrt{\frac{\rho_{e_\mathrm{x} e_\mathrm{y}}}{\alpha_{\rho}}} \right),
    \qquad
    \alpha_{\rho}\geq\max_{e_\mathrm{x}, e_\mathrm{y}}\rho_{e_\mathrm{x} e_\mathrm{y}},
    \label{eq:rho-controlled-rotation}
\end{equation}
on an additional ancilla $a_{\rho}$ followed by uncomputation $\mathbf{O}_{\rho}^\dagger$, gives a block-encoding of $\sqrt{\mathbf{D}_{\rho}}$ with a normalization constant $\sqrt{\alpha_{\rho}}$. 
When the spatially varying Young's modulus is given as a function of spatial coordinate, another way to implement the block-encoding of $\sqrt{\mathbf{D}_{\rho}}$ is to use the diagonal block-encoding technique of spatial coordinate and apply the polynomial transformation to the coordinate~\cite{rosenkranz2025quantum, yano2026quantum}.

If $E_{e_\mathrm{x} e_\mathrm{y}}$ is generated by an efficiently computable structured representation, the oracle may be constant or polylogarithmic in $N_\mathrm{el}$. 
On the other hand, loading an arbitrary matrix $E_{e_\mathrm{x} e_\mathrm{y}}$ can dominate the total complexity. 
If the Poisson ratio also varies by element, the same fixed diagonalizer $\mathbf{V}$ remains valid for the square isotropic element family, while the eigenvalue rotations in Eq.~\eqref{eq:multiplexed-ry-angle} become controlled by the element-wise effective Poisson ratio. 
Variable geometry or element aspect ratio generally requires a different local construction.

\section{Estimation of the condition number}
\label{app:kappa-estimation}

\setcounter{equation}{0}

To determine the polynomial degree of QSVT, we need to estimate the parameter $\kappa_B$ or its lower bound $\kappa_B^\star$ satisfying Eq.~\eqref{eq:kappa-B-FEM}.
From Eq.~\eqref{eq:kappa-lower-bound}, we have
\begin{equation}
  \frac{1 + \nu}
    {1 - \nu / 3} \kappa(\mathbf{K}_{\mathrm F})
    \geq \kappa_B^\star,
  \label{eq:app-kappa-requirement}
\end{equation}
Therefore, an upper bound on the condition number of the stiffness matrix provides a valid choice of $\kappa_B$.
In general, however, the condition number is not known a priori. 
Here, we consider a heuristic approach to estimating the condition number by extrapolation from smaller problems with the same aspect ratio, material properties, and boundary conditions, whose condition numbers can be computed classically.

Motivated by the asymptotical scaling $\kappa (\mathbf{K}_\mathrm{F}) \in \Theta(h^{-2})$ in two-dimensional problems, we employ the following empirical model, including a first-order correction
\begin{equation}
    \kappa (\mathbf{K}_\mathrm{F}) \simeq \widehat{\kappa} = c_0 h^{-2} + c_1 h^{-1},
  \label{eq:app-kappa-fit}
  \end{equation}
where $\widehat{\kappa}$ is the estimated condition number.
The coefficients $c_0$ and $c_1$ are determined by least-squares fitting.

As an example, consider two-dimensional domain $\Omega =(0,a) \times (0,1)$ with $a=1$ or $4$, discretized by square Q1 elements with size $h$.
Displacement is fixed at the left boundary.
We set $\nu=0.3$ and $\lambda=1$.

We first fit the model using the calibration meshes $h\in\{1/4,1/8,1/12,1/16\}$ and evaluate its accuracy on the validation meshes $\mathcal{H}=\{1/20,1/24,1/28,1/32\}$.
Based on the observed underestimation, we introduce a safety factor
\begin{equation}
  s = 1.10 \max \left\{1, \max_{h \in \mathcal{H}}
    \frac{\kappa}{\widehat{\kappa}}\right\},
  \label{eq:app-kappa-margin}
\end{equation}
where the factor $1.10$ is an empirically chosen margin.

We then refit the model using only the validation meshes $\mathcal{H}$ to predict the condition numbers on fines meshes $h=1/64$, $1/96$, and $1/128$, since finer meshes are expected to better capture the asymptotic scaling of the condition number.
The parameter used for QSVT is chosen as
\begin{align}
    \kappa_B
    =
    \frac{1+\nu}{1-\nu/3}
    s\widehat{\kappa}(h).
    \label{eq:kappa_B_est}
\end{align}

Table~\ref{tab:app-kappa-results} summarizes the estimation results.
For all tested meshes, the estimated parameter satisfies $\kappa_B\geq\kappa_B^\star$.
Nevertheless, this extrapolation procedure is heuristic and does not guarantee the inequality for arbitrary mesh sizes or problem settings.
A rigorous guarantee would require a certified upper bound on the condition number.

\begin{table}[t]
  \centering
  \caption{Results of condition number estimation.
  The estimation error is
  $100(\widehat{\kappa}/\kappa-1)$ [\%].
  $N_y = 1/h$ and $N_x = a N_y$.}
  \label{tab:app-kappa-results}
  \begin{tabular}{lll}
    \hline
    $N_x\times N_y$  & Error [\%]
      & $\kappa_B / \kappa_B^\star$ \\
    \hline
    $64\times64$ & $0.149$ & $2.45$ \\
    $96\times96$ & $0.225$ & $2.45$ \\
    $128\times128$ & $0.269$ & $2.45$ \\
    $256\times64$ & $0.094$ & $2.45$ \\
    $384\times96$ & $0.142$ & $2.45$ \\
    $512\times128$ & $0.169$ & $2.45$ \\
    \hline
  \end{tabular}
\end{table}

\end{appendices}

\bibliography{sn-bibliography}%

\end{document}